\documentclass[preprint]{aastex}

\shorttitle{The Distribution of Star-forming Regions in LIRGs}
\shortauthors{Hattori et al.}
\begin{document}

\title{A Study of the Distribution of Star-Forming Regions in
Luminous Infrared Galaxies by Means of H$\alpha$ Imaging
Observations}

\author{
  T. Hattori,\altaffilmark{1,2,3}
  M. Yoshida,\altaffilmark{4}
  H. Ohtani,\altaffilmark{1,2}
  H. Sugai,\altaffilmark{1,2}
  T. Ishigaki,\altaffilmark{1,5}
  M. Sasaki,\altaffilmark{1,6}
  T. Hayashi,\altaffilmark{1,7}
  S. Ozaki,\altaffilmark{1,8}
  M. Ishii,\altaffilmark{1,9}
  and
  A. Kawai\altaffilmark{1,2}
}
\altaffiltext{1}{Visiting Astronomer, Okayama Astrophysical Observatory
of National Astronomical Observatory.}
\altaffiltext{2}{Department of Astronomy, Faculty of Science, Kyoto
  University, Sakyo-ku, Kyoto 606-8502, Japan}
\altaffiltext{3}{Present address: Okayama Astrophysical Observatory,
  National Astronomical Observatory of Japan, Kamogata-cho, Asakuchi-gun,
  Okayama 719-0232, Japan; hattori@oao.nao.ac.jp}
\altaffiltext{4}{Okayama Astrophysical Observatory, National Astronomical
  Observatory of Japan, Kamogata-cho, Asakuchi-gun, Okayama 719-0232, Japan}
\altaffiltext{5}{Department of Applied Physics, Graduate School
  of Engineering, Hokkaido University, Kita-ku, Sapporo, Hokkaido
  060-8628, Japan}
\altaffiltext{6}{Shimonoseki City University, Shimonoseki, Yamaguchi
  751-8510, Japan}
\altaffiltext{7}{Toyama Science Museum, Toyama, Toyama 934-8084, Japan}
\altaffiltext{8}{Nishi-Harima Astronomical Observatory, Sayo-cho,
  Hyogo 679-5313, Japan}
\altaffiltext{9}{Kurashiki Science Center, Kurashiki, Okayama 712-8046, Japan}

\begin{abstract}

We performed H$\alpha$ imaging observations of 22 luminous infrared
galaxies to investigate how the distribution of star-forming regions in these galaxies is related to galaxy interactions.
Based on correlation diagrams between H$\alpha$ flux and continuum
emission for individual galaxies,
a sequence for the distribution of star-forming regions was found:
very compact ($\sim$ 100~pc) nuclear starbursts with almost no star-forming
activity in the outer regions (type~1),
dominant nuclear starbursts $\lesssim$ 1~kpc in size and a negligible
contribution from the outer regions (type~2),
nuclear starbursts $\gtrsim$ 1~kpc in size and a significant
contribution from the outer regions (type~3),
and extended starbursts with relatively faint nuclei (type~4).
These classes of star-forming region were found to be strongly related to global star-forming
properties such as star-formation efficiency, far-infrared color, and
dust extinction. There was a clear tendency for the objects with more compact
distributions of star-forming regions to show a higher star-formation efficiency
and hotter far-infrared color.
An appreciable fraction of the sample objects
were dominated by extended starbursts (type~4), which is unexpected in
the standard scenario of interaction-induced starburst galaxies.
We also found that the distribution of star-forming regions was
weakly but clearly related to galaxy morphology:
severely disturbed objects had a more concentrated
distribution of star-forming regions.
This suggests that the properties of galaxy interactions, such as dynamical
phase and orbital parameters, play a more important role
than the internal properties of progenitor galaxies, such as dynamical structure
or gas mass fraction.
We also discuss the evolution of the distribution of star-forming regions in interacting
galaxies.

\end{abstract}


\section{Introduction}

Many statistical studies of interacting galaxies have shown that galaxy interactions can enhance star-forming activity
\citep[e.g.,][]{larson78,kennicutt87,hummel87,bushouse88,xu91}.
Luminous infrared galaxies (LIRGs; $L_{\rm IR}$[8--1000 \micron]
$\geq$ $10^{11}$ $L_\odot$) are galaxies which emit the bulk of
their energy in the far-infrared (FIR) and often show signs of
interaction, such as tidal tails, multiple nuclei, or disturbed
outer envelopes \citep{sanders96}.
LIRGs are considered to be extreme objects,
where strong starbursts are induced by galaxy 
interactions, because in many such objects star 
formation can account for the infrared emission.

Theoretical work supports the idea that interactions play an
important role in stimulating starbursts in galaxies.
Numerical simulations of merging gas-rich spiral galaxies show
that during the merging process, gas clouds lose their angular
momentum and flow into the circumnuclear region of the host
galaxies \citep[e.g.,][]{barnes91,mihos96}.
The resulting high concentration of molecular gas can fuel star-formation activity in the circumnuclear region.
This is consistent with compact nuclear starbursts and gas
condensation in ultra-luminous infrared galaxies (ULIRGs;
$L_{\rm IR}$ $\geq$ $10^{12}$ $L_\odot$), as observed in mid-infrared
\citep[MIR,][]{soifer00}, radio continuum
\citep[hereafter C91]{condon91}, and CO emission
\citep[e.g.,][]{downes98,bryant99} data. Therefore, a nuclear starburst
triggered by gas inflow has generally been assumed to be the
mechanism for producing enhanced star formation in interacting galaxies.

However, the diversity of star-forming and morphological properties
in LIRGs, and the details of the physical processes that may
enhance star-forming activity, are not fully understood.
For example, several paired systems of LIRGs show intense star formation
in the region of overlap of the two galaxies
\citep{xu00,mirabel98,haas00,soifer01,lefloch02}, which
is not predicted by numerical simulations.
Moreover, some observational studies have suggested that the
distribution of star-forming regions in LIRGs depends on
pair separation \citep[hereafter H99]{hwang99}, gas content
\citep{mihos98}, or degree of morphological
disturbance \citep{xu00}.
Although these observations yield intriguing clues, the samples are
small (4--8 objects) and 
the mechanism that determines the distribution of star-forming
regions in LIRGs remains unclear.
As for the morphological properties, observational studies of
$IRAS$ galaxies have revealed a weak dependence of global star-forming
properties, such as infrared luminosity and FIR color, on the
interaction strength or morphological class \citep[e.g., ][]{xu91,lutz92}.
These studies suggest that in order to better understand the physical
processes in interaction-induced star-forming activity it is
necessary to determine how the star-forming properties of interacting
galaxies depend on the dynamical or physical properties of the system.
On the other hand, even if objects with similar $L_{\rm IR}$ or FIR
color are considered, there is a significant variation in morphological
appearance.
For example, 20--30\% of LIRGs with 10$^{11}$ $L_\odot$ $<$ $L_{\rm IR}$ $<$
10$^{12}$ $L_\odot$ are apparently single galaxies, while the remaining objects
are closely interacting or merging galaxies \citep{sanders96}.
It is unclear why these galaxies are actively forming stars at the same level as
strongly interacting galaxies, and whether the single-galaxy LIRGs have nuclear starbursts.

The distribution of star-forming regions is expected to be connected
to the process by which star-formation activity is enhanced
\citep[e.g.,][]{mihos98}, and therefore it is
important to investigate the distribution of star-forming regions in LIRGs
and its dependence on morphological, physical, and dynamical properties.
For this purpose, we have undertaken an H$\alpha$ imaging survey of LIRGs.
In contrast to previous H$\alpha$ imaging surveys of interacting galaxies
\citep{kennicutt87,bushouse87,combes94,domingue03} and infrared-selected
galaxies \citep[hereafter AHM; \citealt{lehnert96,xu00,dopita02}]{armus90},
all of our sample objects are
LIRGs, and they are not selected based on morphological properties or FIR
color. This should enable us to focus on activated star
formation, and may be useful for examining the relationships between the
distribution of star-forming regions and other properties.
As LIRGs are dust-rich systems, the H$\alpha$ emission, especially the
total H$\alpha$ flux \citep{dopita02}, can be significantly affected by
dust extinction.
However, \citet{lehnert96} presented evidence that the half-light radius
of H$\alpha$ emission is a useful indicator of the size of the star-forming
region. Therefore, H$\alpha$ emission may be used to trace the spatial distribution
of star-forming regions within individual galaxies.
An advantage of
H$\alpha$ emission is its large range of spatial scales,
i.e., it is sensitive to small ($\sim$ 1\arcsec) and large
($\sim$ 1\arcmin) scale emission.
Although observations in MIR or radio continuum emission can provide the
highest spatial resolutions ($\sim$ 0\farcs1 or smaller), such observations
are insensitive to diffuse emission \citep[C91;][]{soifer00,soifer01},
which can extend well beyond several arcseconds.
Continuum images also offer important information,
since red images taken through a tunable filter are almost
free from emission lines and can therefore be used to trace the
distribution of old stars.
This enables a direct comparison of the distribution of old and
newly-formed stars.

In \S~2, we present
the sample selection and the observational details.
In \S~3, we describe the results of the observations for the individual objects
and analyses of the resulting dataset.
The observational results are discussed in \S~4, and the
conclusions and summary are given in the final section.

\section{Observation}

\subsection{Sample Selection}
\label{section_sample}

The sample objects were selected from LIRGs observed in CO emission to
enable comparison of our results with the gas content and star-formation efficiency.
Among the systematic CO surveys of LIRGs
\citep[e.g.,][hereafter SSS]{mirabel90,sanders91},
our sample objects were selected from SSS, where total CO fluxes
were obtained for 62\% of the LIRGs in the {\it Infrared Astronomical
Satellite} ($IRAS$) Bright Galaxy Sample
\citep[BGS;][]{soifer89,sanders95}.
The BGS includes galaxies covering nearly the entire sky at galactic
latitudes higher than 5$^\circ$ with $F_\nu$(60~\micron) $\geq$ 5.24 Jy.
The SSS selection criteria result in a sample that is relatively unbiased to any galaxy property.
Uncertainties in the CO fluxes are typically smaller than 20\%.

We selected objects from the SSS sample matching the following criteria:
(1) $10^{11}$ $L_\odot$ $\leq$ $L_{\rm IR}$ $\leq$ $10^{12}$ $L_\odot$,
(2) no clear evidence for an active galactic nucleus (AGN),
(3) $cz$ $\leq$ 12000 km s$^{-1}$, and
(4) $\delta$ $\geq$ $-10^\circ$.
No morphological selection was made.
The narrow range in infrared luminosity was chosen to avoid confusion due to the effects of large differences in the classes of star-formation activity, as there is a weak dependence of the half-light radius of H$\alpha$
emission on $L_{\rm IR}$ \citep[AHM;][]{lehnert96}.
This should aid investigation of the relation
between properties of the interacting systems and the distribution
of star-forming regions.
In addition, LIRGs with $L_{\rm IR}$ $\leq$ $10^{12}$ $L_\odot$
are relatively close as compared to ULIRGs, making it easier to
investigate the distribution of star-forming regions in detail.
We searched for evidence for AGN in observational studies that used
optical and MIR spectroscopy, radio morphology, and hard X-ray emission.
Based on these criteria, a total of 28 objects was selected, and
observations were made of 22 of them.
Among the 22 objects,
some (III~Zw~35, NGC~2623) show LINER-like optical
spectra \citep{veilleux95}.
However, their radio morphology, large FIR/radio
and small $f_{25}/f_{60}$ ratios suggest that star-forming activity
dominates the energetics of these objects. Hence, the LINER spectra may
result from a contribution due to shock heating, possibly driven by
superwind activities \citep{taniguchi99,lutz99}. For all of the sample
objects, the FIR/radio ratio is similar to or larger than that of normal
star-forming galaxies and fairly constant, with a standard deviation
of $\sim$ 0.14 in $\log L_{\rm FIR}/L_{\rm 1.4GHz}$. This suggests that
none of the sample objects harbors an energetically dominant AGN.
Several objects in our sample are in common with those studied in previous
H$\alpha$+[\ion{N}{2}] imaging surveys of LIRGs; four with AHM
(NGC~1614, NGC~2623, NGC~4194, and UGC~8335),
one with \citet[Zw~049.057]{lehnert95}, one with
\citet[NGC~6621/22]{xu00}, and two with
\citet[NGC~1614 and NGC~3110]{dopita02}.
The relatively small number of common objects is due to differences in the
adopted selection criteria \citep[AHM;][]{lehnert95,xu00} or to
different sky coverage \citep{dopita02}.
The observed objects are listed in Table~\ref{obs_list} together
with their global parameters: gas mass, infrared luminosity, and
FIR fluxes at 25, 60, and 100 \micron \ determined by SSS,
\citet{soifer89}, and \citet{sanders95}.

Most of the sample objects are interacting systems, and many of
them contain two or more galaxies in the $IRAS$ beam; the FIR
fluxes may therefore represent the total emission from the component galaxies,
and so it is possible that there is no LIRG in the system
\citep[see e.g.,][]{sulentic93}.
To examine this possibility, we searched for radio continuum observations of
objects that have nearby companions or are composed of two galaxies.
These objects are listed in Table~\ref{pair_list},
with $L_{\rm IR}$ for each component galaxy determined by assuming that they both have the same
radio-to-FIR flux.
For Mrk~331, no radio information was found for the companion galaxies.
However, as the 100 \micron \ mapping observation of Mrk~331 by
\citet{zink00} found a flux consistent with the $IRAS$ BGS flux, contamination by
companion galaxies is considered to be negligible.
Of the ten objects listed in Table~\ref{pair_list}, the first five are
paired galaxies included in $IRAS$ BGS as interacting systems because
of their small separations.
The infrared luminosities of these
objects were found to be dominated by one of the member galaxies, and
so the analyses in the following sections are made for the dominant galaxy only.
The H$\alpha$ luminosity can also be used to resolve the confusion problem
because it is closely correlated to FIR luminosity for normal star-forming
galaxies \citep[e.g.,][]{kewley02}.
However, in LIRGs, the $L_{\rm IR}/L_{\rm H\alpha}$ ratio varies over an
order of magnitude (AHM). In contrast, if there is not an energetically
dominant AGN, the FIR/radio ratio is fairly constant, as is the case in
our sample. Therefore, we did not use the H$\alpha$ luminosity for this purpose.

\subsection{H$\alpha$ and Continuum Imaging}
\label{section_imaging}

The H$\alpha$ and continuum images were obtained in 1999 December,
2000 January and October, and 2001 April using the f/18 Cassegrain focus
of the Okayama Astrophysical Observatory (OAO) 188-cm telescope.
We used the Kyoto Tridimensional Spectrograph I \citep{ohtani98} in imaging Fabry--Perot interferometer mode with a 5 \micron \ gap
Queensgate ET-50 etalon. This etalon provides bandpasses of
25 \AA \ at 6598 \AA \ and 29 \AA \ at 6929 \AA. A 4\farcm5 square
field is imaged onto a 1024$\times$1024 TC215 CCD with a pixel size
of 12 \micron. By 2$\times$2 on-chip binning, each 24 \micron \ pixel
subtends 0\farcs535 on the sky.
Observational procedures for the Kyoto Tridimensional Spectrograph I
and the system setup are described in detail in \citet{ishigaki}.

The central transmission wavelength of the etalon was tuned to the
redshifted H$\alpha$ emission of each object. Images of the continuum
emission were also obtained at the rest-frame wavelengths of 6450~\AA \
and/or 6650~\AA.
A log of these observations is presented in Table~\ref{obs_log},
which lists the observed wavelength, the exposure time, and
the full width at half maximum (FWHM) of field stars.
The contribution of H$\alpha$ + [\ion{N}{2}] to the continuum image is
generally small, estimated to be about 1--10\% for the typically
observed H$\alpha$ equivalent widths (EWs) of 10--100\AA.
Sixteen of our sample objects were observed by \citet{kim95} and
their results show that the nuclear [\ion{N}{2}]/H$\alpha$ ratio ranges
from 0.3 to 1.0.
This means that [\ion{N}{2}]
emission contributes up to 30\% of the obtained H$\alpha$ images.
A clear example may be III~Zw~35, which shows, in addition to compact
nuclear emission, elongation in the direction of the minor axis, suggesting
that galactic-scale outflows may contribute to the H$\alpha$ image.
However, the compact distribution of star-forming regions in III~Zw~35 is
not greatly affected by the elongated component, and in the following analyses we consider such
effects to be generally small.

The images were reduced in the standard manner for
narrow-band filter imaging using IRAF.
The procedure includes bias subtraction, dark subtraction,
flat-fielding, cosmic-ray removal and sky subtraction.
Flux calibrations were made for most of the objects
by using data from photometric standard
stars taken with the same instrumental setup,
although
relatively unstable weather conditions caused rather
large uncertainties in the calibration.
A comparison of the integrated fluxes with previous H$\alpha$ imaging
observations \citep[AHM;][]{lehnert95,dopita02} in six of the 
sample objects showed log differences of about 0.3, and we therefore
assumed this value as the uncertainty in the absolute flux
calibration.
No flux calibration was made for NGC~828, NGC~6090, and Mrk~331.
Spatial registration and correction for the seeing variation were
made using the positions and FWHM of field stars.
Where reliable corrections for variations of system efficiency
with wavelength and atmospheric transmission were not available,
the counts of field stars were used to match the flux level of
on- and off-band images. As the off-band images were obtained
at both the long- and short-wavelength sides of the H$\alpha$ emission,
the effect of the color difference between field stars is expected to be small.

\section{Results}

\subsection{H$\alpha$ and Continuum Images}
\label{section_hacon_image}

The continuum-subtracted H$\alpha$ and continuum images of the sample
objects are shown in Figure~\ref{ha_con_image}.
By taking the ratio of the H$\alpha$ and continuum images, we obtained
the H$\alpha$ equivalent-width maps,
which are also given in Figure~\ref{ha_con_image}.
For objects with an H$\alpha$-emitting companion galaxy,
the images of the companion galaxies are shown separately in
Figure~\ref{companion_image}.
The images show a large variation in the distribution of H$\alpha$
emissions, and analyses of the H$\alpha$ images are made in the next
subsection.
A description of the results for each individual object is given in the appendix.

In continuum light,
the sample objects span a broad range of morphologies from apparently normal spiral galaxies (e.g., NGC~958) to nearly complete
mergers with strong tidal tails (e.g., NGC~2623).
It is notable that in systems of paired galaxies most of the
companion galaxies are small as compared to the main galaxies that are
the dominant source of FIR emission.

We divided the sample objects into several morphological
classes in order to facilitate comparisons with the star-forming properties given in the following
sections.
Morphological classifications have already been made by SSS
for all of our samples.
However, according to their classification, more than half
of the galaxies are defined as "merger", partly because their "merger" class includes paired
galaxies with overlapping disks.
Therefore,
we made our own morphological classification as follows.
Objects with strong tidal features and a single nucleus were classified
as "merger" (NGC~1614, NGC~2623, Arp~148, NGC~4194).
Paired galaxies with overlapping disks or a connecting bridge were
classified as "close pair" (III~Zw~35, Arp~55, UGC~8335, NGC~6090,
NGC~6621/22) where the projected separation ranges from $\sim$ 4~kpc to
20~kpc.
Galaxies which have nearby ($<$ 100~kpc) companion galaxies at the same
redshift but no evident overlap between the primary galaxy and
its companion were classified as "pair" (NGC~877, NGC~3110, IC~2810,
NGC~7771, Mrk~331). 
The remaining eight objects were classified as "single".
Note that, although the "single" objects do not show strong tidal
features, many of them show some distorted/asymmetric appearance.
Thus, it is possible that some of the "single" objects are remnants
of mergers or have experienced tidal interactions.
In Figure~\ref{ha_con_image},
the morphological classes are indicated to the left of the H$\alpha$
images, where "M", "C", "P", and "S"
represent "merger", "close pair", "pair", and "single", respectively.

\subsection{Distribution of H$\alpha$ Emission and Equivalent Width}
\label{section_ha_con_diagram}

As seen in Figure~\ref{ha_con_image}, the morphological appearance of
LIRGs in H$\alpha$ is complicated and highly varied.
To investigate the distribution of star-forming regions and compare it
with physical and dynamical properties,
we made "H$\alpha$--continuum diagrams."
These diagrams plot the H$\alpha$ versus red-continuum surface brightness
at each galaxy position, and were made for each individual object.
The surface brightness in red continuum light, which is considered
to trace old stars, typically decreases with distance from the nucleus, 
so the diagram demonstrates how the H$\alpha$ surface brightness changes with distance
from the nucleus and, more specifically, the relative concentration towards the nucleus of H$\alpha$ and continuum emission.
Note that for paired systems, only the main galaxy, that emits most of
the FIR flux (see Table~\ref{pair_list}), is shown in the diagram.
The use of continuum surface brightness as an indicator of the distance
from the galaxy center has some advantages, in that it is insensitive to variations in galaxy size and inclination; however, there may be additional
uncertainty due to non-axisymmetric features such as bars and spiral arms.
In normal spiral galaxies, the distribution of H$\alpha$ emission is
similar to that of white light \citep{hodge83}, and more extended
than $I$-band emission \citep{ryder94}.

The H$\alpha$--continuum diagrams are shown in Figure~\ref{ha_cont} for
all of the objects, where the ordinate and abscissa represent log H$\alpha$
and continuum surface brightness in erg s$^{-1}$ cm$^{-2}$ arcsec$^{-2}$
and erg s$^-1$ cm$^{-2}$ \AA $^{-1}$ arcsec$^{-2}$.
For NGC~828, NGC~6090, and Mrk~331, the axes are shown in arbitrary units,
as we did not make absolute flux calibrations for these objects.
However, in these three
objects, the relationship between the ordinate and abscissa is the same as
in other objects, and H$\alpha$-equivalent widths can be measured.
The abscissa increases leftward, so the galaxy nucleus corresponds
to the left edge of the diagram.
Before making these diagrams, the spatial sampling and FWHM of the
point spread function were set to common values of 420 pc pixel $^{-1}$
and 1.43 kpc, respectively, for all of the objects.
In the diagrams, small dots represent the H$\alpha$ and continuum
surface brightnesses measured at each pixel of the image.
If the distribution of H$\alpha$ and continuum emission are similar,
the small dots will be distributed along lines with a
slope of 1.0, as shown in Figure~\ref{ha_cont}.
These three lines correspond to a constant EW of 20~\AA, 100~\AA,
and 500~\AA, respectively.
In the upper panel of each diagram, we show the H$\alpha$ flux integrated
over the regions with the continuum surface brightness given by the
abscissa value.
Each continuum surface-brightness bin has a width of 0.1 in logarithmic
scale, and the ordinate is linearly scaled, with the bottom line
corresponding to zero H$\alpha$ flux.
In these plots, one can see the contributions to the total H$\alpha$ flux
from the nuclear and outer regions.
Hereafter, they are referred to as "H$\alpha$-flux histograms".

The diagrams in Figure~\ref{ha_cont} are ordered from upper left to lower
right by characteristics of appearance, and we classified them as follows.
The H$\alpha$-flux histograms of the first eight objects (III~Zw~35,
NGC~2623, Zw~049.057, NGC~1614, NGC~4194, UGC~8335, NGC~6090, and Mrk~331)
show that the total H$\alpha$ flux of these objects is dominated by
emission from the nuclear region.
The H$\alpha$--continuum diagrams
show a monotonic decline of H$\alpha$ surface brightness towards the right-hand side, and the
position of the H$\alpha$ peak closely matches that of the continuum peak.
In addition, most of the slopes are steeper than 1.0,
i.e., the H$\alpha$ distribution is more sharply peaked than the continuum.
In spite of these common properties, there are two subgroups with distinctly
different H$\alpha$ surface brightnesses and EW: III~Zw~35, NGC~2623, and
Zw~049.057 show log $f_{\rm H\alpha}$ $\sim$ $-15$ (in cgs) and H$\alpha$
EW lower than 20\AA in their nuclei, while the other five objects show log
$f_{\rm H\alpha}$ $\sim$ $-13.5$ and H$\alpha$ EW larger than 100\AA.
Hereafter, the former group is referred to as type~1 and the latter as
type~2.
In the next six objects (NGC~23, UGC~2238, Arp~55, IC~2810, NGC~6621,
and NGC~7771), the H$\alpha$-flux histograms clearly show that there are
significant contributions from the outer regions to the total H$\alpha$ flux.
In addition, although the H$\alpha$ peaks are located near the nuclei, the
spread of the plotted points at the left edge of the H$\alpha$--continuum
diagrams indicates that the H$\alpha$ peaks are displaced from the continuum
peaks.
We call this class of objects type~3.
The next seven objects (NGC~695, NGC~828, NGC~834, NGC~877, NGC~958,
NGC~3110, and NGC~5653) are galaxies in which most of the H$\alpha$ emission
comes from the outer regions. Many of them have the strongest H$\alpha$ emission
far from the nucleus and show a more extended distribution of H$\alpha$
than of continuum emission.
These objects are type~4.
Arp~148 does not match any of the types described above. This may be due to
the apparent double components in continuum light, probably caused by heavy dust
extinction at the nucleus, which is located halfway between the components (see
appendix~\ref{section_a148}),
or it may be related to the peculiar appearance of the large off-center ring.
In any case, Arp~148 is not included in any of the above types and
is treated as an exceptional object.

To parameterize the distribution of H$\alpha$ emission, we also
calculated the peak surface brightness to total flux ratio ($P/T$) and
the H$\alpha$-emitting area
($S_{\rm H\alpha}$) in the spatial-resolution and -sampling adjusted
images.
$P/T$ measures the relative importance of the nuclear region to the
disk region in star-forming activity and was used by H99
as a measure of the concentration of MIR emission.
$S_{\rm H\alpha}$ is defined as the total area of emission-line region(s)
brighter than a threshold value of surface brightness.
We adopt a value of 3.5 $\times$ 10$^{-16}$ erg s$^{-1}$ cm$^{-2}$
arcsec$^{-2}$ for the threshold, which is well above the background
noise for most of the images.
In practice, $S_{\rm H\alpha}$ can be used to distinguish whether
star-forming activity is confined to the nuclear region or extends to
the outer regions.
The global properties and the parameters of the H$\alpha$ distribution
are listed in Table~\ref{global_prop},
and Table~\ref{global_differ_table} lists the mean and standard
deviation of these values for each type.
In these tables, $C_{60\mu}$ is the spectral curvature
at 60 \micron \, defined by C91 as the difference in the
spectral indices, $C_{60\mu} \equiv \alpha_{25\mu,60\mu} -
\alpha_{60\mu,100\mu}$.

\subsubsection{Star-forming Properties of Each Type}
\label{section_type_property}

From Table \ref{global_differ_table},
it is clear that the classification based on the
diagrams in Figure~\ref{ha_cont}
is associated with the physical characteristics of star-forming
activity.
For example, the star-formation efficiency (SFE), measured by
$L_{\rm IR}/M_{\rm H_2}$ \citep{young86}, and the dust temperature,
estimated from the $f_{60}/f_{100}$ ratio, are significantly higher
in type~1 and type~2 than type~3 and type~4 objects.
In addition, type~1 objects generally show exceptionally large or small
values in $S_{\rm H\alpha}$, $M_{\rm H_2}$, $L_{\rm IR}/L_{\rm H\alpha}$,
and $C_{60\mu}$ relative to the other types.

More evidence for differences between the types of galaxy is given by the
$\log (f_{60}/f_{100})$--$\log (f_{25}/f_{60})$,
$C_{60\mu}$--$\log (L_{\rm IR}/L_{\rm H\alpha})$,
and $\log(P/T)$--$\log (L_{\rm IR}/L_{\rm H\alpha})$ diagrams
shown in Figure~\ref{global_differ}.
These diagrams illustrate that different types
occupy different parts of the diagrams.
The diagrams also indicate that objects of the same type have
surprisingly similar star-forming properties.
In the following sections, we summarize the observational properties
of each type
and attempt to reveal the nature of star-forming activity in them.

Figure~\ref{global_differ}b shows a positive correlation
between $C_{60\mu}$ and $\log (L_{\rm IR}/L_{\rm H\alpha})$ with a
correlation coefficient of 0.88.
As shown by C91, LIRGs
also show a definite positive correlation between $C_{60\mu}$ and
"hybrid surface brightness", which was defined as the ratio of the total
60 \micron \ flux and the angular area in 8.44~GHz or 1.49~GHz emission
of an object.
The correlation shows that objects with large $C_{60\mu}$ tend to have
a compact radio source, and therefore a large hybrid surface brightness.
C91 concluded that the large values of $C_{60\mu}$ in compact radio
sources result from large optical thicknesses at $\lambda$ $\leq$ 25 \micron.
In these sources, the H$\alpha$ emission may also suffer from strong
dust extinction, resulting in the large $L_{\rm IR}/L_{\rm H\alpha}$ values.
Therefore, the correlation in Figure~\ref{global_differ}b suggests that
the variations in $C_{60\mu}$ and $\log (L_{\rm IR}/L_{\rm H\alpha})$ in our
sample objects are dominated by the variations in the amount of dust
extinction.
If this is the case, we note that in Figure~\ref{global_differ}b the
type~1 objects suffer the strongest dust extinction and the type~4 objects
suffer the weakest.

\subsubsection{Type~1 and Type~2}
\label{section_type12}

The type~1 and type~2 objects are characterized by warm dust temperatures
(large $f_{60}/f_{100}$),
large SFE ($L_{\rm IR}/M_{\rm H_2}$), and strongly concentrated H$\alpha$
emission.
This means
that these galaxies are forming stars actively in their nuclear regions.

In addition to the above properties, the type~1 objects show extreme
properties also in other observational quantities, i.e., they have the
largest $L_{\rm IR}$/$L_{\rm H\alpha}$ and $C_{60\mu}$,
and the smallest $S_{\rm H\alpha}$ and $M_{\rm H_2}$ among the four
types.
The small $S_{\rm H\alpha}$ indicates that the star-forming
activity in these objects is confined to their nuclei, and little
star formation occurs in the outer regions, as is clearly seen in the
original images of Figure~\ref{ha_con_image}b, j, r.
The large $C_{60\mu}$ suggests that these objects have compact nuclear
star-forming regions, as noted in \S~\ref{section_type_property}.
Actually, NGC~2623 and III~Zw~35, two of the type~1 objects, were
observed by C91 and described as "dominated by compact radio components."
The physical size of the compact components is about 100~pc.
In Zw~049.057, another type~1 object, the radio continuum emission at
1.49~GHz is dominated by a nuclear source with an estimated diameter
of 0\farcs5 \citep{condon90}, which corresponds to $\sim$ 100~pc.
The strong dust extinction for nuclear star-forming regions, as
deduced from the large $L_{\rm IR}/L_{\rm H\alpha}$ and $C_{60\mu}$, may have the effect of decreasing the observed H$\alpha$ emission, causing the faint H$\alpha$ surface brightnesses
and small EW in these objects.
As for the gas content,
the mean value of $M_{\rm H_2}$ is a factor of two smaller than for the other
types, although dispersions are large.

The H$\alpha$-flux histograms of type~2 objects are similar
to those for type~1 objects.
However, the
morphologies observed in the original images clearly differ:
while the H$\alpha$ and continuum images of type~2 objects are
similar in the nuclear region, those of type~1 objects
are not.
Moreover,
high resolution radio/MIR observations reveal the following difference in the
distribution of star-forming regions between the two types.
UGC~8335 was observed by C91 and clearly shows an extended structure
with a size of $\sim$ 1\arcsec \ ($\sim$ 600 pc), in contrast to
III~Zw~35 and NGC~2623, which are dominated by compact radio components
as noted above.
In addition, the star-forming regions in NGC~1614, NGC~4194, NGC~6090,
and Mrk~331 are known to be spatially extended to several hundred parsecs
or 1~kpc
\citep{soifer01,aalto00}.
Thus, it would appear that although both type~1 and type~2 objects show dominant nuclear
starbursts, the nuclear star-forming regions are
slightly larger in type~2 objects than in type~1 objects.
The bright H$\alpha$ surface brightness and the large EW
in type~2 objects may be the result of a strong concentration of star-forming regions
and
relatively small dust extinction, as inferred from small
$L_{\rm IR}/L_{\rm H\alpha}$.

\subsubsection{Type~3}
\label{section_type3}

As seen in the H$\alpha$--continuum diagrams, objects of this type
show active star formation in their nuclear region.
However, there is a substantial contribution from the outer regions to
the total H$\alpha$ flux.
In addition, the dust temperature and SFE are lower than for the two
types above.

Examination of the original H$\alpha$ images revealed that all of the
type~3 objects have bright outer star-forming regions that are extended
on scales of 10~kpc. The significant contributions from the outer regions seen
in the H$\alpha$-flux histograms originate in these star-forming regions.
These are associated with the disks (NGC~23, UGC~2238,
IC~2810, and NGC~7771), the tidal tail (Arp~55), or the overlap
region (NGC~6621).
In addition, they all show slight offsets of the order of 1~kpc
between the H$\alpha$ and continuum peaks.
The offsets are not artificial effects due to incorrect subtraction of
continuum emission,
because most of them show significant differences in their nuclear
surface-brightness profiles even between the on- and off-band images, i.e.,
before the subtraction of continuum emission.
In NGC~7771, there is a circumnuclear starburst ring with a diameter of
$\sim$ 1.5~kpc \citep{smith99,reunanen00}.
Although the ring cannot be seen in our H$\alpha$ image, probably because of
the coarser spatial resolution, the offset between the H$\alpha$ and continuum
peaks in NGC~7771 is clearly due to the presence of the starburst ring
(see appendix~\ref{section_n7771}).
This suggests that the nuclear star-forming regions in the other type~3
objects are also extended on scales of 1~kpc or larger.
If this is the case, the size of the nuclear star-forming region in these
objects is similar to or larger than the adjusted point spread function, having a FWHM of 1.43~kpc.
That the nuclear star-forming regions are slightly larger in size than the point
spread function naturally explains why,
in spite of the nearly identical range of star-formation rates ($L_{\rm IR}$),
the type~3 objects have fainter nuclear H$\alpha$ surface brightnesses
(typically by a factor of ten) and smaller EW than type~1 objects.
The reduced peak surface brightness and the large contribution from
the outer regions to the total H$\alpha$ flux may cause
the smaller values of $P/T$ as compared to type~1 and type~2 objects
(Figure~\ref{global_differ}c).

\subsubsection{Type~4}
\label{section_type4}

As noted in \S~\ref{section_ha_con_diagram},
the total H$\alpha$ fluxes of the type~4 objects are dominated by emission
from extended star-forming regions.
They have small $P/T$, low dust temperatures, and small SFE
and $L_{\rm IR}$ (Table~\ref{global_differ_table}).
These properties indicate that
the star-forming activity in these objects is relatively mild
as compared with the other types.
It is probably because of this that observations of these objects have been very limited in the literature to date.

A remarkable characteristic of the distribution of star-forming regions
in the type~4 objects is their knotty and asymmetric appearance.
In particular, NGC~834 and NGC~5653 have a dominant extra-nuclear
H$\alpha$ source, which results in large values of $P/T$
As compared to the other type~4 objects (see Figure~\ref{global_differ}c).
It is notable that all of the type~4 objects show small values of
$L_{\rm IR}/L_{\rm H\alpha}$ (Figure~\ref{global_differ}b,c),
suggesting that the amount of dust extinction in these objects
is relatively small.
Thus, it is unlikely that the relative faintness of their nuclei
in H$\alpha$ is due to suppression by strong dust extinction.
We examined this question in more detail by searching for images that trace
the star-forming regions and are relatively insensitive to dust extinction.
As noted in appendix~\ref{section_n5653}, the Pa$\alpha$ image of NGC~5653
\citep{alonso02} shows features that are nearly identical to our H$\alpha$ image.
In Figure~\ref{condon}, we compared the distribution of H$\alpha$ in
NGC~695 and NGC~3110 with 1.49-GHz radio continuum maps \citep{condon90}
with matched spatial resolution of $\sim$ 5\arcsec.
Apart from a slight offset of the peak position in NGC~695,
the overall appearance in H$\alpha$ and radio continuum emission is
similar in the two galaxies.
Moreover, the radio emission extends beyond 10~kpc, and the
total flux is dominated by emission from the outer regions as
in H$\alpha$.
In this respect, they are distinctly different from the other types,
most of which show dominant nuclear sources in radio continuum emission.
Therefore, the star-forming activity in the type~4 objects is probably
dominated by an extended starburst rather than a nuclear one.

The type~4 objects account for an appreciable fraction of our
60 \micron \ flux-limited sample of LIRGs: 32\% in the whole sample
and 46\% in the lower luminosity range,
$L_{\rm IR}$ $\leq$ $10^{11.5}$ $L_\odot$.
This suggests that, in addition to nuclear starbursts, star formation
extending to several kiloparsecs is another attribute
of star-forming activity in LIRGs.

\section{Discussion}

\subsection{Distribution and Global Properties of Star Formation}
\label{section_distribution_property}

In \S~\ref{section_ha_con_diagram}, it is shown that the size of the nuclear
star-forming region
increases from type~1, through type~2, to type~3 objects.
In the type~3 objects,
there are substantial contributions from the outer regions to the total
H$\alpha$ flux.
The star-forming activity in the type~4 objects is dominated by extended
starbursts.
If the distribution of molecular gas clouds is similar to that of
star-forming regions,
the density of gas clouds may also vary from type to type, being highest
in the type~1 objects and lowest in type~4 objects.
The higher density of gas clouds in galaxies with compact star formation
will lead to a higher SFE, because of the non-linear dependence of the star-formation
rate on gas density \citep{kennicutt98}.
The high efficiency of star formation will then produce more
photons to heat the dust and, consequently, raise the dust temperature
\citep{young86}.
In Figure~\ref{global_differ}a, galaxies with a strong concentration
(type~1 and type~2) show the largest values of $f_{60}/f_{100}$,
i.e., the highest dust temperature, whereas
galaxies that are dominated by extended starbursts (type~4) show the
smallest values, with type~3 lying between the two extremes.
Except for type~1, the $f_{25}/f_{60}$ ratio shows a similar tendency
to $f_{60}/f_{100}$, although the difference between the types is less clear.
The small values of $f_{25}/f_{60}$ in the type~1 objects are also
expected to be caused by
the compact distribution
of nuclear star-forming regions,
as discussed in \S~\ref{section_type12}.
Therefore, the difference in the distribution of star-forming regions
between the types can explain their segregation in
Figure~\ref{global_differ}a.

In addition to the compactness of the nuclear star-forming regions
($\sim$ 100 pc),
the type~1 objects are remarkable in
the small amount of molecular gas present and
the nearly complete lack of star-forming activity in the outer regions.
These properties suggest that they are in a late stage in the
star-formation history of LIRGs \citep{gao99}, and that most of the
remaining gas has fallen to the nuclear region.
In addition, all of the type~1 objects are either OH megamaser sources
or OH absorbers.
This suggests that there is a strong radio source in the type~1 objects,
located behind an extended molecular-gas screen.
It is notable that, although LIRGs with $L_{\rm IR}/L_{\rm H\alpha}$ as
large as type~1 are rare, Arp~220, the prototypical ULIRG, shows comparably
faint H$\alpha$ emission and large $L_{\rm IR}/L_{\rm H\alpha}$ ratio (AHM).
The FIR color and the presence of an OH megamaser source in Arp~220 are also
similar to type~1 objects.
Thus,
it is probable that, as in Arp~220, the energy source of type~1 objects
is a compact starburst deeply embedded in dusty molecular clouds
\citep[see also \citealt{baan87}]{downes98,sakamoto99}.
This would cause the large $L_{\rm IR}/L_{\rm H\alpha}$ and the small
$f_{25}/f_{60}$ seen in type~1 objects.

From the above discussion, we conclude that differences in the
distribution of star-forming regions causes the differences in 
star-forming properties among the types of object.
Therefore, the H$\alpha$--continuum diagram combined with the
H$\alpha$-flux histogram (Figure~\ref{ha_cont}) is a useful tool for
investigating the properties of star-forming activity in LIRGs.

\subsection{Morphological Consideration}
\label{section_morph_type}

In this subsection, we investigate the relationship between
star-forming properties and the morphological class of the sample
objects.
As the object types defined in \S~\ref{section_ha_con_diagram} were found to
be strongly correlated with the distribution and global properties of
star formation, it may be useful
to compare our type classification with the morphological class of the galaxies.

In Figure~\ref{morph_type_hist}, histograms of the types
are shown for each morphological class.
The type~1 and type~2 objects are relatively small in number
and are both dominated by nuclear starbursts, so the objects of the two
types are combined and labeled as "Type 1/2" in
Figure~\ref{morph_type_hist}.
The morphological classes are defined in \S~\ref{section_hacon_image}. Dynamical disturbance due to galaxy interactions is expected to be
the strongest in
"mergers" and to gradually become weaker in the order "close pairs", "pairs", and
"singles".
Therefore, Figure~\ref{morph_type_hist} implies that
the more severely disturbed objects
tend to have a more compact distribution of star-forming regions.
Although there are several exceptions to this rule, we suggest
that the degree of dynamical disturbance plays an important role in
determining the distribution of star-forming regions.

From MIR imaging of seven LIRGs with morphological properties indicating the
early/intermediate stages of merging, H99 found that the
peak-to-total flux ratio, which is a measure of the concentration of
star-forming regions (see \S~\ref{section_ha_con_diagram}), increases
as the projected separation of the interacting pairs becomes smaller.
Morphologically,
their entire sample corresponds to the class of "close pairs" in our classification
scheme, and their FIR colors are similar to our type~2 and type~3 objects.
In addition,
the objects with larger projected separations have FIR
colors similar to type~3 objects, and the objects with smaller projected separations 
have FIR colors similar to type~2 objects.
Therefore, H99's and our results are complementary:
whereas H99 revealed
the detailed dependence of the distribution of star-forming regions
on the interaction stage in a morphologically selected sample, our
results show that the dependence can be extended to a sample with
a broader range of morphological properties.
H99 interpreted their results as a manifestation of the evolutionary
sequence: as the merging process advances, star formation becomes
more active and develops a more concentrated distribution.
A similar interpretation may also be applicable to our sample, with the exception of the
single galaxies. This issue will be discussed in the next subsection.

\citet{mihos98} and \citet{xu00} also found connections between
dynamical states and the distribution of star-forming regions.
From imaging Fabry--Perot observations of four ULIRGs, \citet{mihos98}
found that the H$\alpha$ emission is more spatially concentrated in the
later stages of a galaxy interaction.
\citet{xu00} found that relatively undisturbed systems
show evenly distributed MIR and H$\alpha$ emission in the disks of
component galaxies, while severely disturbed objects show dominant
nuclear components and enhanced emission in the overlap regions of the
pairs.
As the star-forming activity of their sample objects
differs significantly from ours, we do
not make detailed comparisons with these studies.
Nevertheless, it is notable that these studies obtained qualitatively
similar results to ours.
Therefore, it is almost certain that the strength of dynamical disturbance
plays a critical role in determining the distribution of star-forming
regions.
Our sample size,
the restricted range of star-formation rates, and the broad range of
morphological properties of our sample objects explain why such a clear
relation emerges in Figure~\ref{morph_type_hist}.

With respect to global star-forming properties, several statistical studies of $IRAS$ galaxies, covering a wide range of morphological properties, have shown that they
are weakly connected to the interaction strength \citep[e.g., SSS;][]{xu91,lutz92}.
Our results suggest that this is because the interaction strength controls
the distribution of star-forming regions: the stronger the
dynamical disturbance, the more compact the region to which the star-forming activity is confined, resulting in the more extreme properties of higher SFE
and higher dust temperature.

\subsection{Evolution of the Distribution of Star-forming Regions}
\label{section_evolution}

In this subsection, we consider the relation between the type and the
star-formation history of LIRGs.

There are remarkable differences in the size of nuclear
star-forming regions and in the relative contribution of the outer regions to
the total H$\alpha$ flux between the different types of galaxy.
These properties, combined with the connection between type and morphological class, suggest a sequence in the
distribution of star-forming regions in interacting galaxies.
At an early stage of interaction, active star formation occurs in both
the nuclear and outer regions (type~3). As the process of interaction
proceeds, the infall of gas clouds activates nuclear star formation,
making it the dominant energy source of the system (type~2).
By the final phase, most of the gas clouds have fallen into the nuclear
region, and the star-forming activity is confined to the nucleus (type~1).
In this process, the size of the nuclear star-forming region also becomes
gradually smaller.

The evolutionary sequence above is similar to that described
by \citet{scoville01}, which invokes star formation induced by cloud-cloud
collisions \citep{scoville86}. In their sequence, the cloud collision rate
is increased by tidally enhanced velocity dispersion in the individual
disks in the initial phase, later by the passage of the disks
through each other, and lastly by concentration of the gas at
smaller radii in the merged nucleus.
Therefore, star formation in the outer regions is expected to be important
at an earlier phase of galaxy interaction, which is consistent with the
observed sequence described above.
As for the variation in size of the nuclear star-forming regions,
a probable cause of the slightly larger ($\gtrsim$ 1 kpc) size in 
type~3 objects is a nuclear resonance ring, such as that observed in NGC~7771
(see \S~\ref{section_type3}).
The circumnuclear star-forming ring might shrink with time as a result of the combined
effect of a non-axisymmetric potential and the self-gravity of gas
\citep{wada92} or due to the friction exerted by background stars on gas
clouds \citep{combes92}.
The timescale of these processes is of the order of $10^8$ years.
As this is shorter than the merger timescale ($\sim$ $10^9$ years),
and in some cases strong gaseous inflows can occur at early stages of galaxy interactions \citep{mihos96},
interacting galaxies in the type~3 phase can evolve into the type~2 or
type~1 phase before the final merger.
Therefore, the absence of a one-to-one correspondence between the
morphological class and the type (Figure~\ref{morph_type_hist}) does not
conflict with this evolutionary sequence.

Since $L_{\rm IR}$ may vary significantly throughout the star-formation history
of LIRGs \citep[e.g.,][]{xu00,murphy01}, type~3 objects may not evolve simply into
type~2 and then into type~1.
Our suggestion is that the distribution of
star-forming regions in interacting galaxies follows the sequence above
and, during some phases, the galaxies are recognized as LIRGs.
Although this evolutionary model seems to succeed in explaining the processes
that occur in the evolution of the distribution of star-forming regions, detailed
studies of individual objects are needed to examine the validity of
these conclusions.

As many of the type~4 objects are singles, we consider them separately.
With the exception of NGC~828,
the star-forming activity in type~4 objects is characterized by either 
prominent spiral arms (NGC~877, NGC~958, NGC~3110) or
the brightest H$\alpha$ source being located in the outer regions
(NGC~695, NGC~834, NGC~5653),
with both of these groups showing a knotty appearance in H$\alpha$.
In addition, their morphological properties are related to the distribution
of star-forming regions. All of the objects with prominent spiral arms
have a small companion at a projected separation of 20--30~kpc, assuming that the faint
galaxy $\sim$ 70\arcsec \ southeast of NGC~958 has the same redshift as
NGC~958. The other objects have peculiar morphologies, with no
galaxy-like object within a projected radius of 100~kpc.
Since strong spiral arms can
be induced by the presence of a small companion \citep[e.g.,][]{toomre81}, this result suggests that the star-forming activity in the first group is enhanced by the presence of such tidally excited spiral arms.
However,
it is unclear whether such a process is efficient enough to induce active
star formation and produce a LIRG.
As for the latter group, in spite of the fact that they appear to be isolated galaxies,
the structural peculiarities and the asymmetric
distribution of star-forming regions suggest that their star-forming
activity is also induced by galaxy interactions.
This result could be explained if they are remnants of either minor
mergers or high speed tidal encounters.
The lack of detailed observational information about the
type~4 objects makes it difficult to examine these possibilities.
Two-dimensional kinematic observations are needed to reveal the nature
of these objects, and may be useful in further developing our understanding of 
star-forming activity in LIRGs.

\section{Summary}

In order to investigate the relation between the distribution of
star-forming regions and galaxy interactions, H$\alpha$ imaging
observations of twenty-two LIRGs were carried out using an imaging
Fabry-Perot interferometer.

\paragraph{The H$\alpha$--continuum diagram}
The distribution of H$\alpha$ emission was investigated with
H$\alpha$--continuum diagrams, using the continuum surface brightness
as an indicator of the distance from the galactic center,
and with H$\alpha$-flux histograms, in which the distribution of
integrated H$\alpha$ flux was shown as a function of continuum
surface brightness.
The sample objects were classified into four
types according to the appearance in the diagrams.
These types were found to be strongly related to global
star-forming properties, such as FIR color, SFE, and dust extinction.
The variations in these properties between the types of object can be explained
by differences in the distribution of star-forming regions.

\paragraph{Connection with morphological properties}
We studied the galaxy morphology of each of the four types, 
and found a clear tendency for severely disturbed objects to show
concentrated distributions of star-forming regions.
This is consistent with previous studies.
However, the larger sample size, broader range of morphological types,
and the restricted range of star-formation rates in our work
provide a clearer result.

\paragraph{Evolution of the distribution of star-forming regions}
From the sequence of the distribution of star-forming regions and its
relation to morphological properties, we propose a scenario in which
the relative importance of nuclear and extended starbursts, as well as
the size of nuclear star-forming regions, changes with time.

\paragraph{Extended starburst}
The star-forming activity in
an appreciable fraction of our 60\micron \ flux-limited sample of LIRGs
is dominated by extended starbursts.
This suggests that star formation extending to several kiloparsecs
is also important in LIRGs.
Therefore,
it may be crucial to determine the triggering mechanism of star formation
in these objects, in order to better understand star-forming activity
in LIRGs.

\bigskip
We would like to thank the staff of Okayama Astrophysical Observatory
for their kind help during the observations.

\clearpage

\epsscale{0.8}
\begin{figure}
 \begin{center}
  \plotone{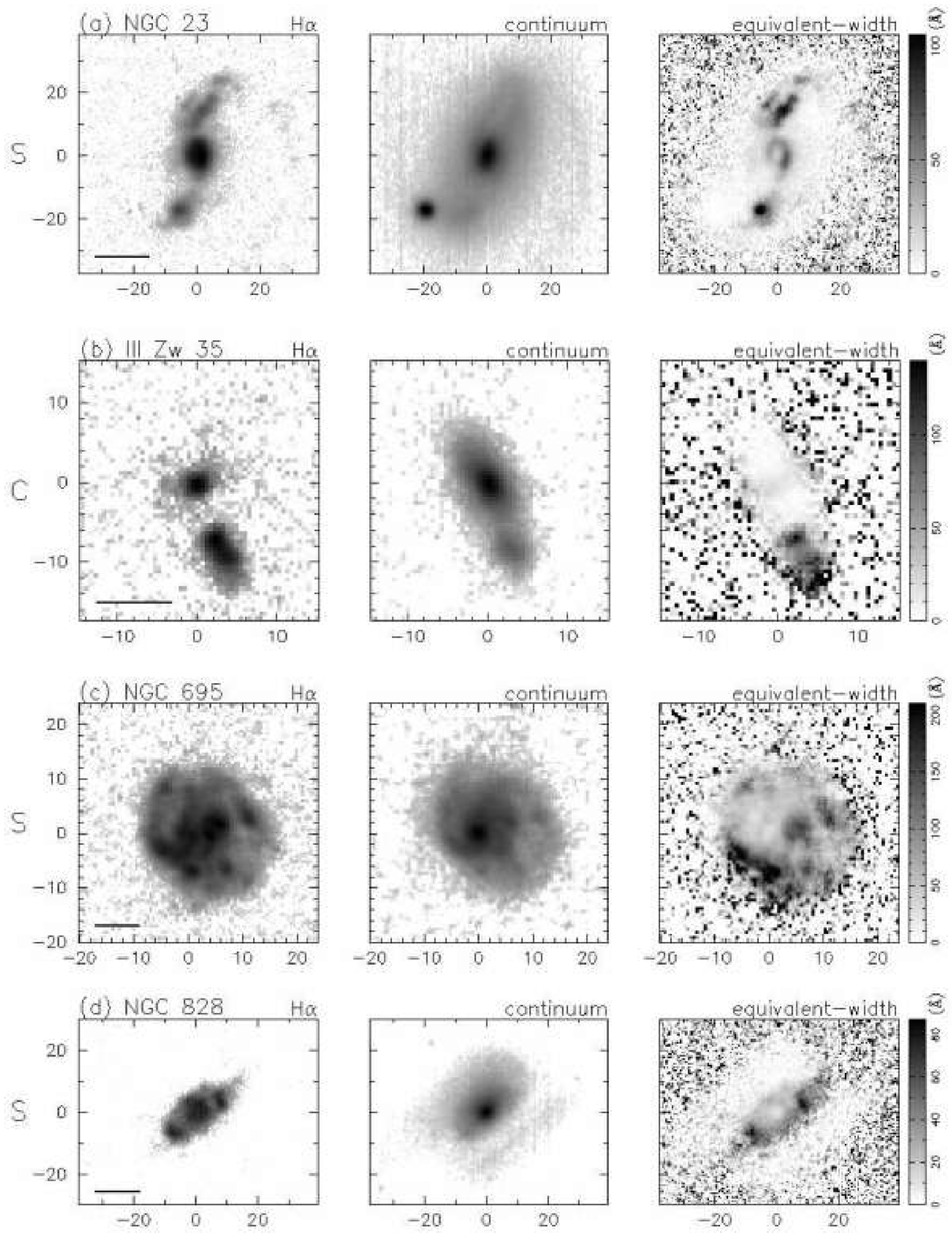}
 \end{center}
 \caption{
  The continuum-subtracted H$\alpha$ images and the continuum images of sample objects.
  The equivalent-width maps are also shown.
  In all cases, north is up, east is to the left, and the coordinates are in units
  of arcseconds.
  The H$\alpha$ and continuum images are shown with logarithmic scales, whereas the
  equivalent-width maps use linear scales as indicated by the scale-bars at
  the right-hand side.
  For each object, the peak position in continuum is set to the origin of the
  coordinates, and a scale of 5~kpc is indicated by a bar located in the
  lower-left corner of the H$\alpha$ image.
 \label{ha_con_image}
 }
\end{figure}

\clearpage
\setcounter{figure}{0}
\begin{figure}
 \begin{center}
  \plotone{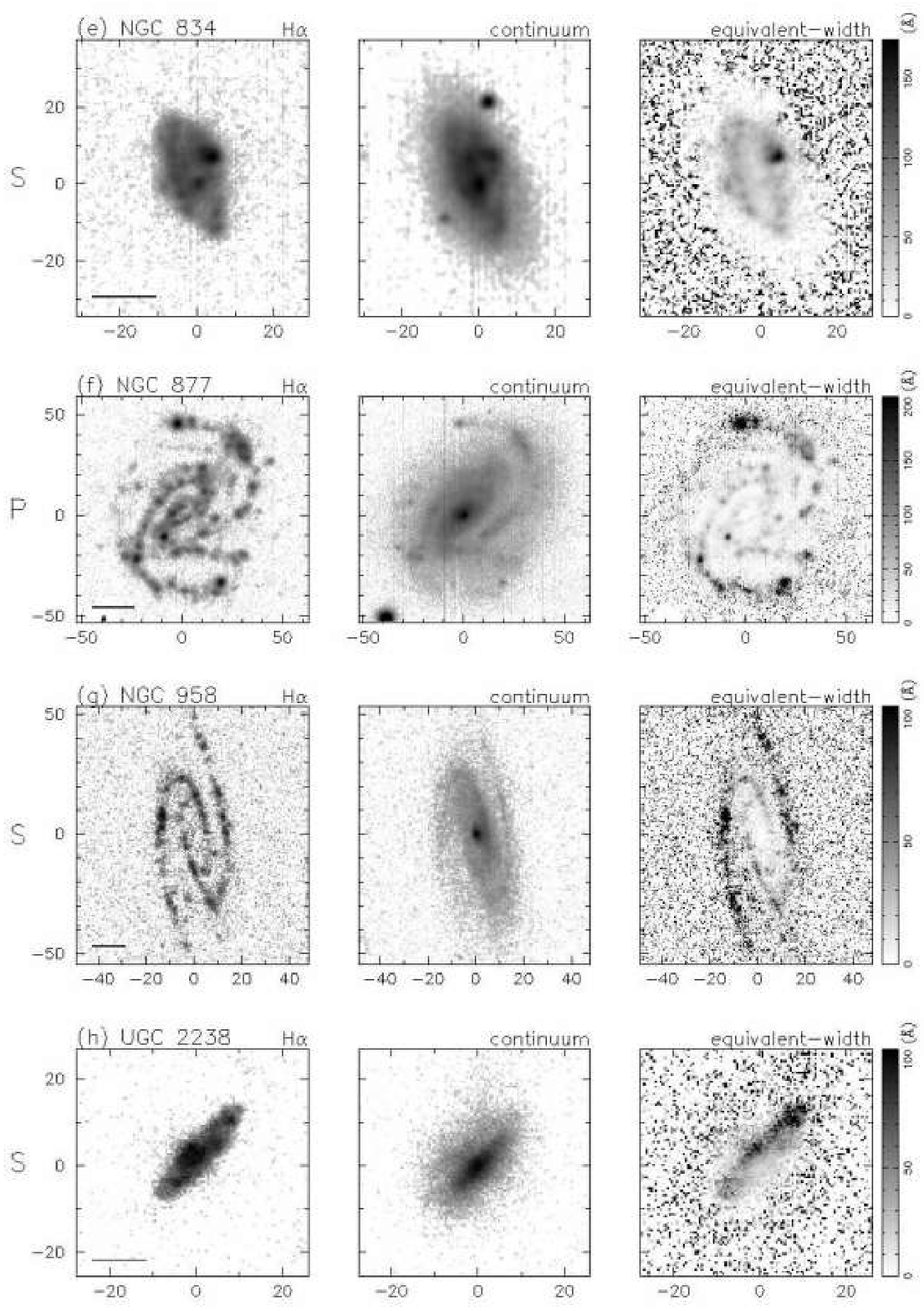}
 \end{center}
 \caption{(Continued)}
\end{figure}

\clearpage
\setcounter{figure}{0}
\begin{figure}
 \begin{center}
  \plotone{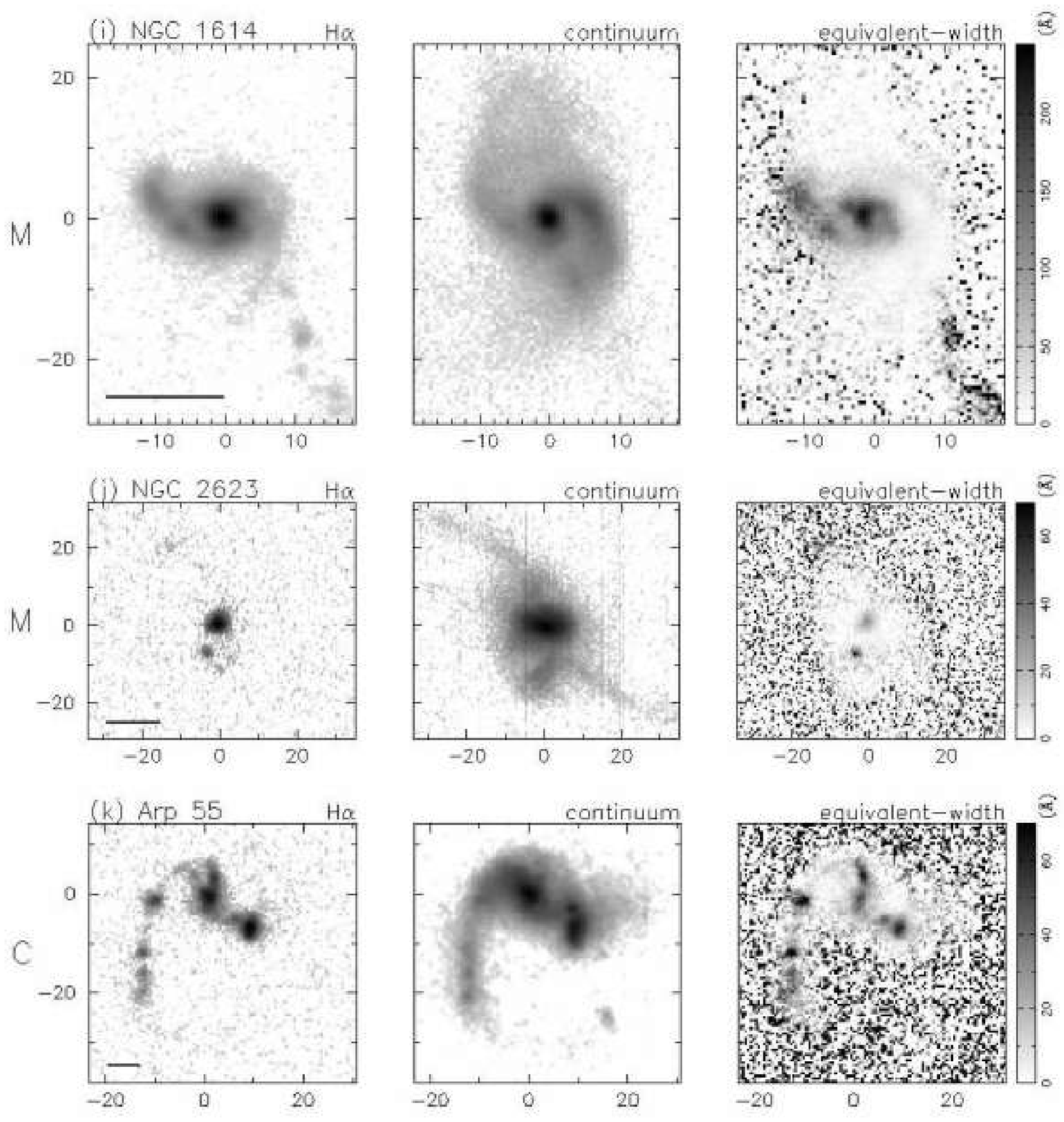}
 \end{center}
 \caption{(Continued)}
\end{figure}

\clearpage
\setcounter{figure}{0}
\begin{figure}
 \begin{center}
  \plotone{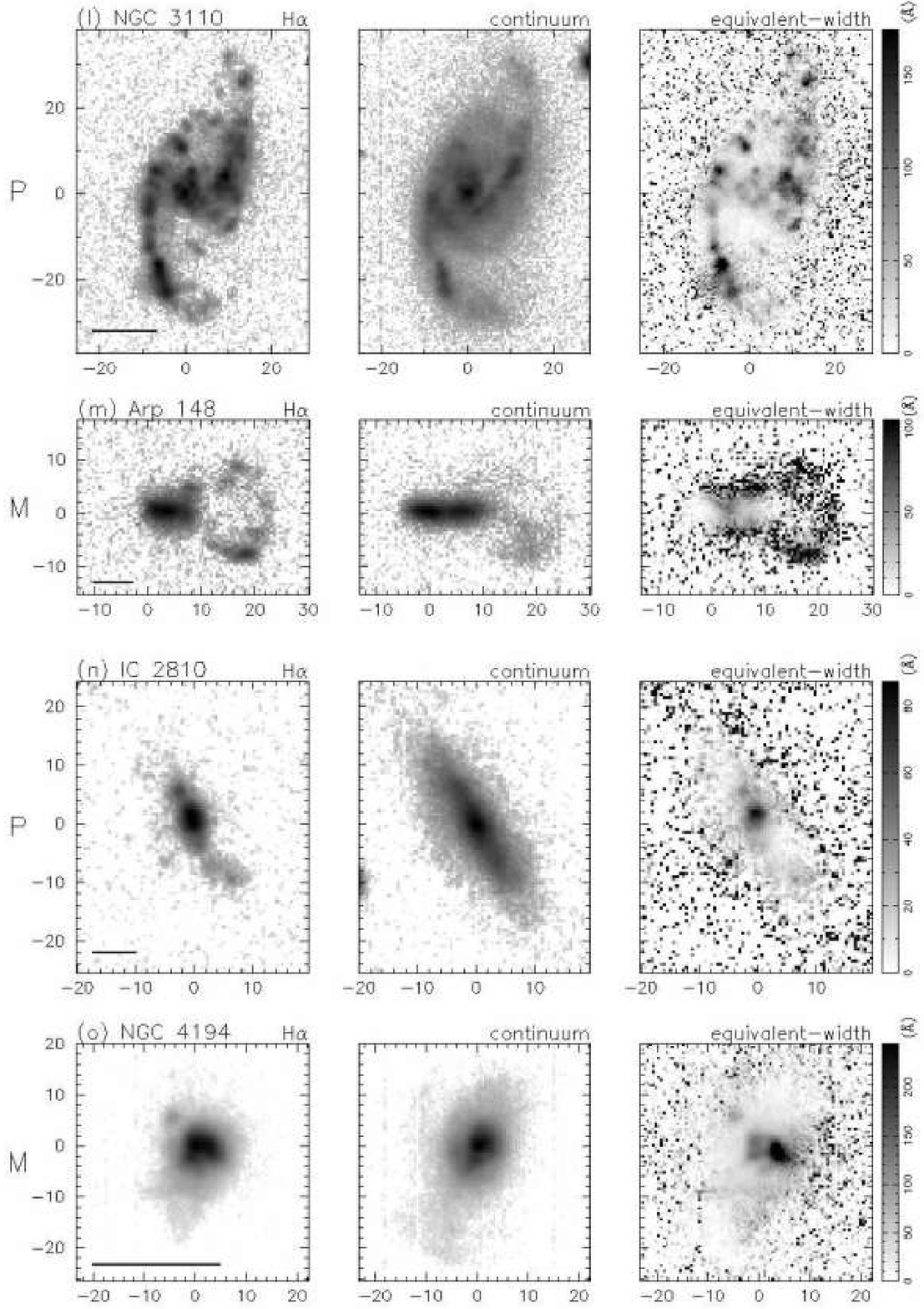}
 \end{center}
 \caption{(Continued)}
\end{figure}

\clearpage
\setcounter{figure}{0}
\begin{figure}
 \begin{center}
  \plotone{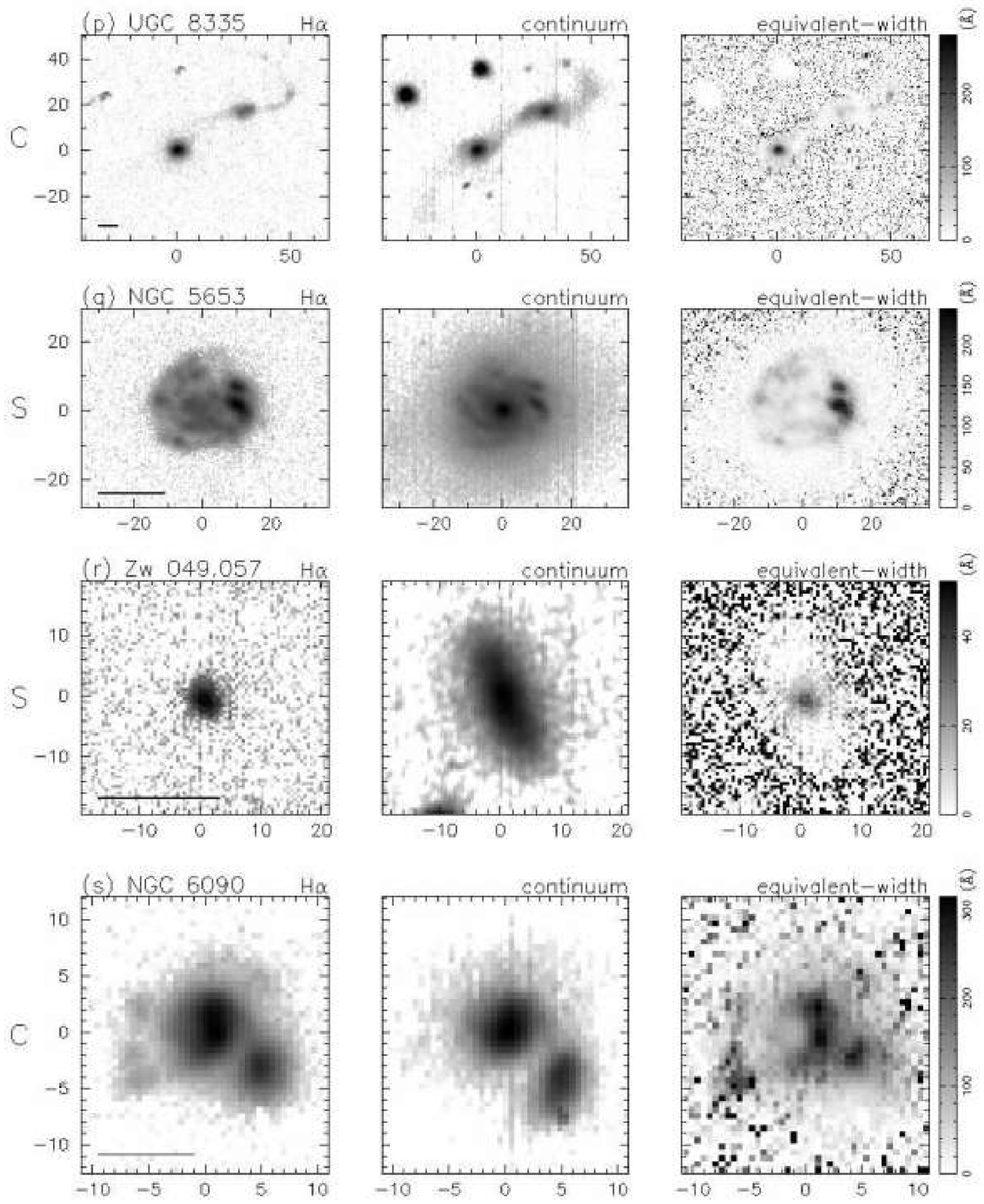}
 \end{center}
 \caption{(Continued)}
\end{figure}

\clearpage
\setcounter{figure}{0}
\begin{figure}
 \begin{center}
  \plotone{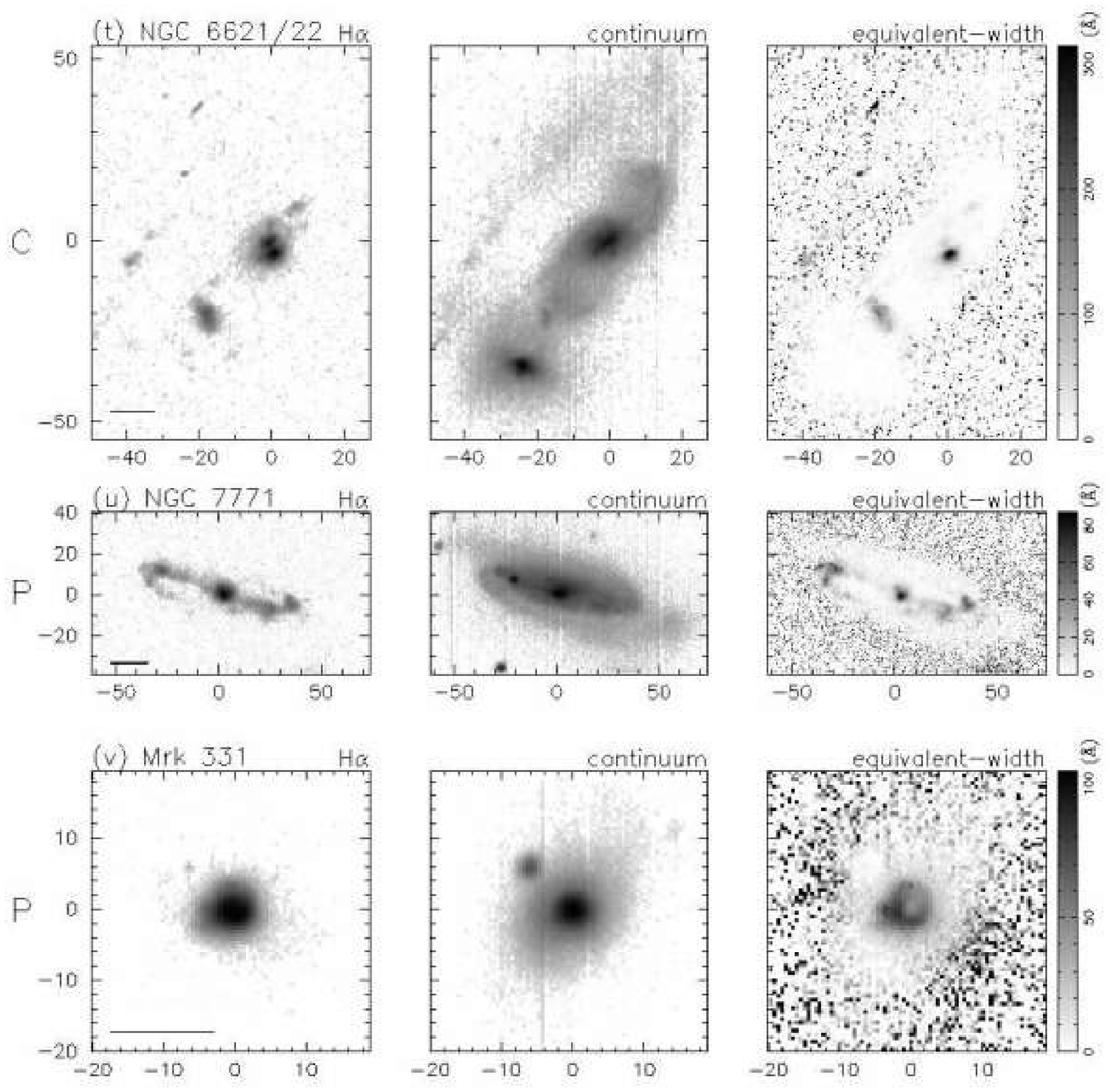}
 \end{center}
 \caption{(Continued)}
\end{figure}

\clearpage
\begin{figure}
 \begin{center}
  \plotone{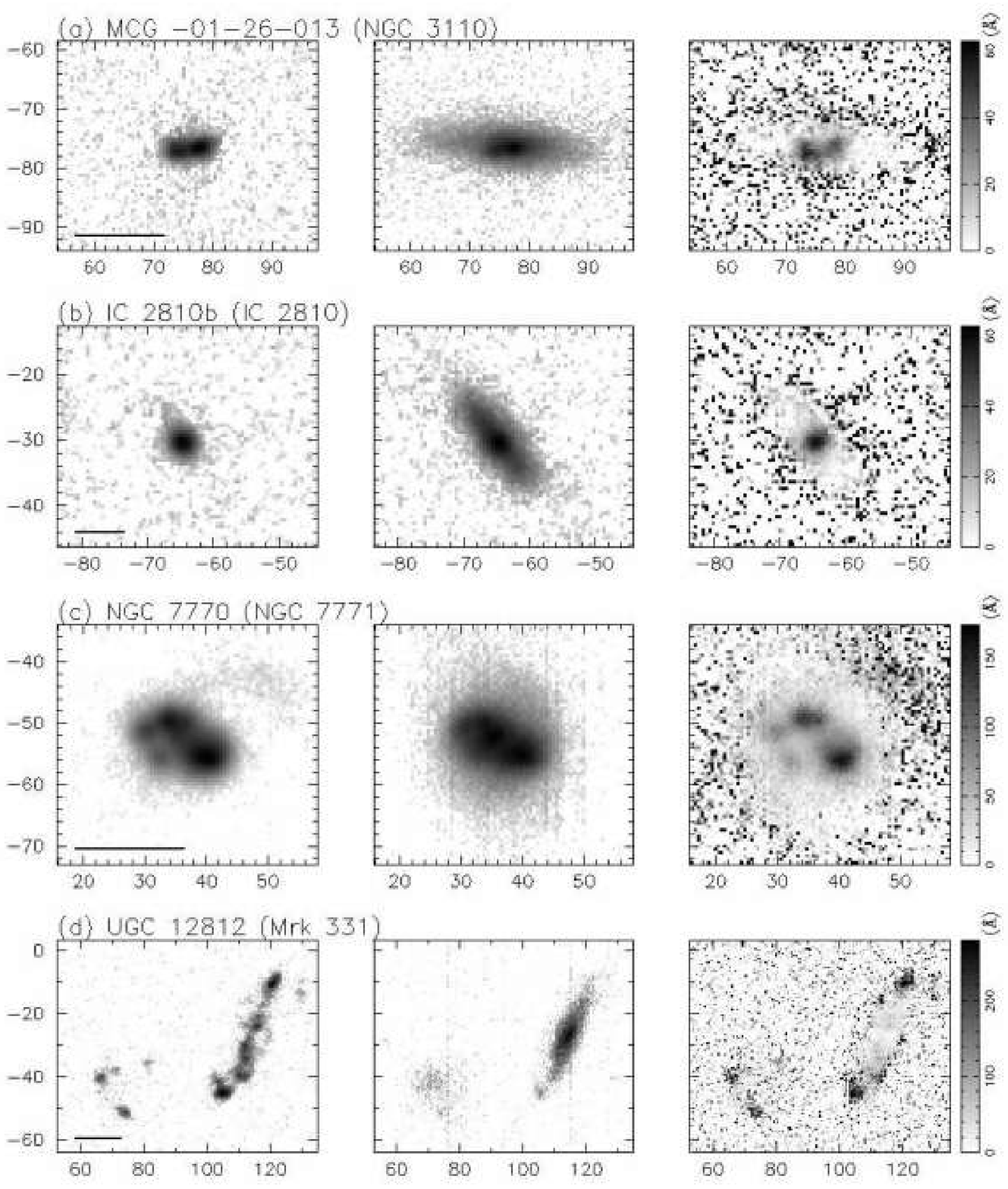}
 \end{center}
 \caption{
  The H$\alpha$ images (left), continuum images (center) and
  equivalent-width maps (right) for the companion galaxies 
  detected in H$\alpha$.
  North is up, east is to the left, and the coordinates are in units
  of arcseconds.
  For each object, the peak position of the main galaxy in continuum
  is set to the origin of the coordinates, and a scale of 5~kpc is
  indicated by a bar located in the lower-left corner of the H$\alpha$
  image.
  The names of the companion galaxies are indicated at the top of the
  H$\alpha$ images and the names of the main galaxies are shown in
  parentheses.
 \label{companion_image}
 }
\end{figure}

\epsscale{0.7}
\clearpage
\begin{figure}
 \begin{center}
 \plotone{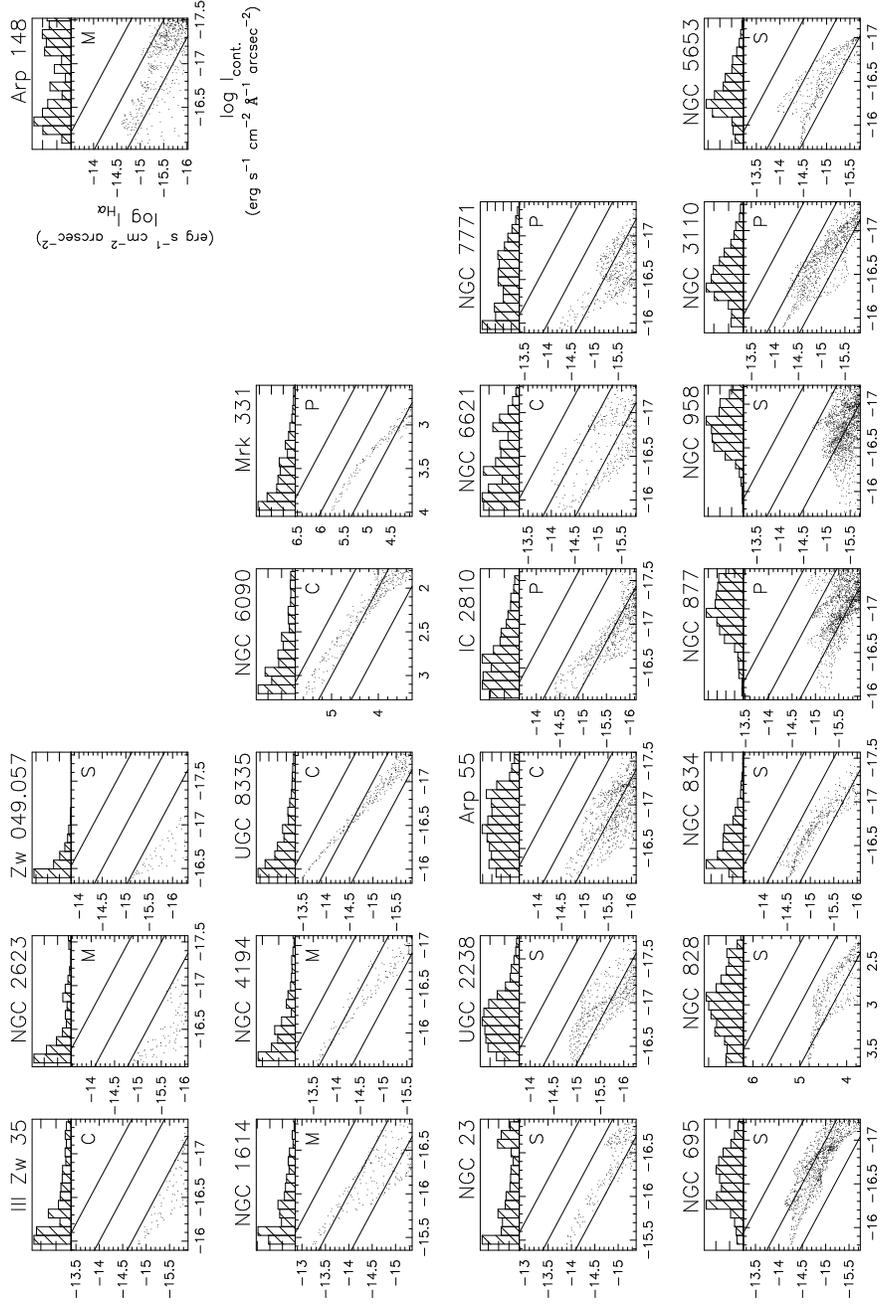}
 \end{center}
 \rotate
 \caption{
  H$\alpha$--continuum diagrams for the sample objects,
  where the ordinate represents log H$\alpha$ surface brightness
  (erg s$^{-1}$ cm$^{-2}$ arcsec$^{-2}$)
  and the abscissa log continuum surface brightness
  (erg s$^-1$ cm$^{-2}$ \AA $^{-1}$ arcsec$^{-2}$).
  Since absolute flux calibrations were not made for NGC~828, NGC~6090,
  and Mrk~331, the axes are shown in arbitrary units for these objects. However, note that in these three objects the relationship between the ordinate and
  abscissa is the same as in the other objects and the H$\alpha$ equivalent widths
  can be measured.
  The continuum surface brightness increases leftward, so the galaxy 
  nucleus corresponds to the left edge of the diagram.
  The small dots represent H$\alpha$ and continuum surface brightness
  measured at each position.
  The three solid lines with slopes of 1.0 correspond to constant
  H$\alpha$ equivalent-widths of 20 \AA, 100 \AA, and 500 \AA, respectively.
  The morphological class is indicated by the symbol in the upper-right
  corner.
  In the upper panel of each diagram, the distribution of integrated
  H$\alpha$ flux from each continuum surface-brightness bin is shown.
  The diagrams are ordered by appearance. See text for details.
 \label{ha_cont}
 }
\end{figure}

\clearpage
\epsscale{0.4}
\begin{figure}
 \begin{center}
 \plotone{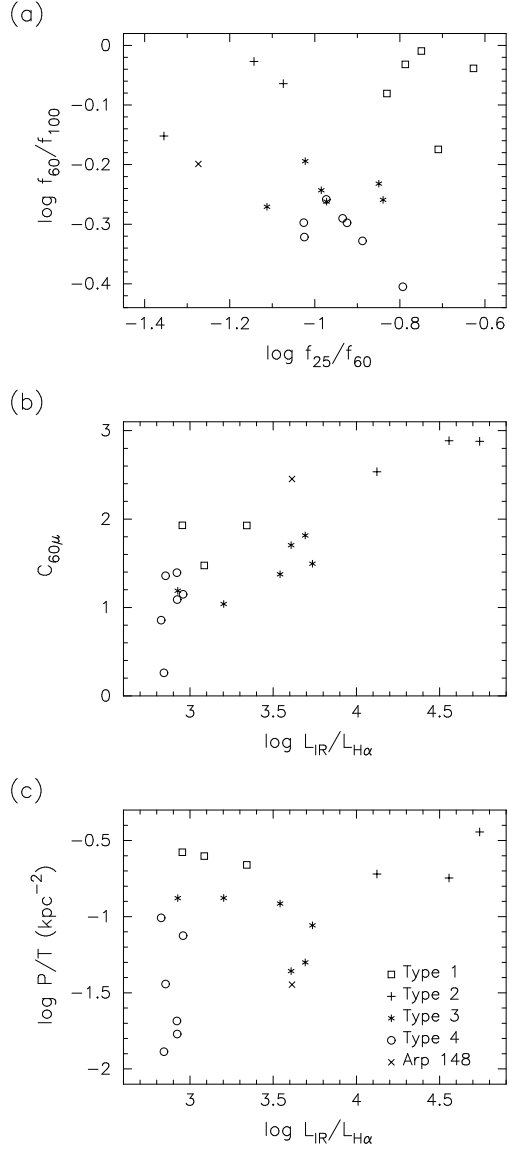}
 \end{center}
 \figcaption{
   Diagrams of $\log (f_{60}/f_{100})$--$\log (f_{25}/f_{60})$ (a),
   $C_{60\mu}$--$\log (L_{\rm IR}/L_{\rm H\alpha})$ (b) and
   $\log (P/T)$--$\log (L_{\rm IR}/L_{\rm H\alpha})$ (c).
   The symbols represent the types of object, as indicated in panel c.
   These diagrams clearly show that the different types of object
   show different global properties.
 \label{global_differ}
 }
\end{figure}

\epsscale{0.8}
\clearpage
\begin{figure}
 \begin{center}
 \plotone{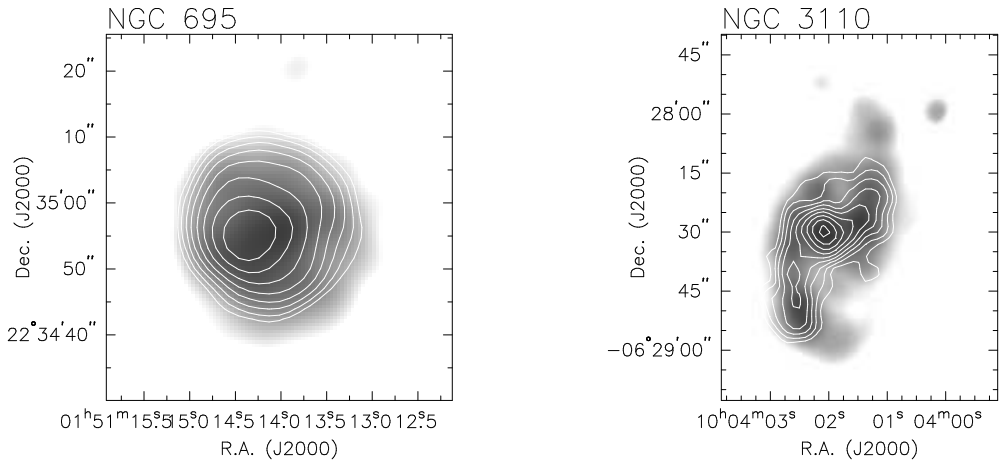}
 \end{center}
 \figcaption{
  Comparison of the distribution of H$\alpha$ and radio continuum
  emission in NGC~695 and NGC~3110.
  The radio data are from \citet{condon90} and the beam size is
  5\arcsec \ at FWHM.
  The H$\alpha$ images were convolved with a Gaussian, so their spatial
  resolution matches that of the radio maps.
  The coordinates in the H$\alpha$ data are determined from field
  stars included in the $Hubble$ $Space$ $Telescope$ Guide Star Catalog,
  and the uncertainty is estimated to be $\lesssim$ 3\arcsec \ from
  comparisons of the radio maps of \citet{condon90} and our H$\alpha$
  or continuum images of several objects.
 \label{condon}
 }
\end{figure}

\epsscale{0.4}
\clearpage
\begin{figure}
 \begin{center}
  \plotone{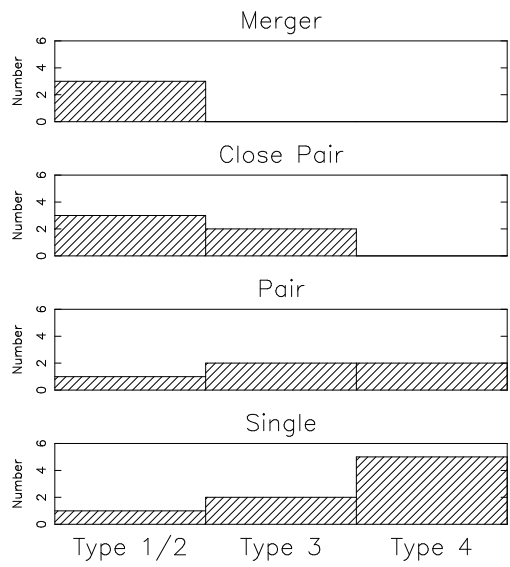}
 \end{center}
 \figcaption{
  Distribution of the types in the H$\alpha$--continuum diagram for
  each morphological class.
 \label{morph_type_hist}
 }
\end{figure}

\clearpage
\begin{deluxetable}{lccccccc}
\tablewidth{0pt}
\tablecaption{List of observed objects.\label{obs_list}}
\tablehead{
\colhead{Name} & \colhead{cz\tablenotemark{a}} &
\colhead{$D$\tablenotemark{a}} &
\colhead{$\log~M_{\rm H_2}$\tablenotemark{a}} &
\colhead{$\log~L_{\rm IR}$\tablenotemark{a}} &
\colhead{$f_{25\micron}$\tablenotemark{b}} &
\colhead{$f_{60\micron}$\tablenotemark{b}} &
\colhead{$f_{100\micron}$\tablenotemark{b}} \\
\colhead{} & \colhead{(km s$^{-1}$)} & \colhead{(Mpc)} &
\colhead{($M_\odot$)} & \colhead{($L_\odot$)} &
\colhead{(Jy)} & \colhead{(Jy)} & \colhead{(Jy)}
}
\startdata
NGC~23 & 4536 & 61.3 & 9.68 & 11.05 & 1.24 & 8.77 & 14.96 \\
III~Zw~35 & 8257 & 109.3 & 9.84 & 11.55 & 1.00 & 11.86 & 13.75 \\
NGC~695 & 9769 & 129.8 & 10.34 & 11.63 & 0.81 & 7.61 & 13.80 \\
NGC~828 & 5359 & 72.6 & 10.14 & 11.32 & 1.04 & 12.03 & 24.38 \\
NGC~834 & 4630 & 62.8 & 9.55 & 10.94 & 0.87 & 6.81 & 12.76 \\
NGC~877 & 3866 & 50.8 & 9.63 & 11.03 & 1.40 & 11.76 & 23.34 \\
NGC~958 & 5896 & 76.4 & 10.14 & 11.16 & 0.95 & 5.90 & 14.99 \\
UGC~2238 & 6478 & 85.2 & 10.14 & 11.27 & 0.63 & 8.16 & 15.22 \\
NGC~1614 & 4847 & 62.3 & 10.03 & 11.61 & 7.82 & 33.12 & 36.19 \\
NGC~2623 & 5538 & 76.1 & 9.77 & 11.54 & 1.85 & 25.72 & 27.36 \\
Arp~55 & 11957 & 162.7 & 10.46 & 11.70 & 0.62 & 6.53 & 10.21 \\
NGC~3110 & 5091 & 69.3 & 10.30 & 11.26 & 1.10 & 11.68 & 23.16 \\
Arp~148 & 10438 & 143.3 & 10.18 & 11.59 & 0.37 & 6.95 & 10.99 \\
IC~2810 & 10299 & 140.5 & 10.26 & 11.53 & 0.58 & 5.60 & 9.80 \\
NGC~4194 & 2620 & 41.4 & 9.28 & 11.09 & 4.57 & 25.66 & 26.21 \\
UGC~8335 & 9296 & 129.2 & 10.13 & 11.76 & 1.96 & 12.01 & 12.92 \\
NGC~5653 & 3618 & 54.4 & 9.63 & 11.03 & 1.33 & 10.27 & 21.86 \\
Zw~049.057 & 3870 & 52.3 & 9.47 & 11.22 & 0.93 & 21.06 & 29.88 \\
NGC~6090 & 8831 & 123.3 & 10.15 & 11.51 & 1.22 & 6.25 & 9.34 \\
NGC~6621/22 & 6234 & 88.4 & 10.01 & 11.22 & 1.02 & 6.99 & 12.35 \\
NGC~7771 & 4346 & 58.6 & 9.95 & 11.35 & 2.18 & 20.46 & 37.42 \\
Mrk~331 & 5385 & 72.3 & 10.11 & 11.41 & 2.56 & 17.32 & 20.86 \\
\enddata

\tablenotetext{a}{These values are taken from SSS.
The $M_{\rm H_2}$ values were calculated from $L_{\rm CO}$ using
a CO to H$_2$ conversion factor of $\alpha$ $=$
3 $\times$ $10^{20}$ cm$^{-2}$ (K km s$^{-1}$)$^{-1}$.
}
\tablenotetext{b}{The $IRAS$ flux densities are from Soifer et al.
(1989) and Sanders et al. (1995).}

\end{deluxetable}

\clearpage
\begin{deluxetable}{lccccc}
\tablewidth{0pt}
\tablecaption{Log of observations.\label{obs_log}}
\tablehead{
\colhead{Name} & \colhead{Date} & \colhead{Wavelength} &
\colhead{Exposure time} & \colhead{Spatial resolution\tablenotemark{a}}
& \colhead{Remark} \\
& & \colhead{(\AA)} & \colhead{(seconds)} & &
}
\startdata
NGC~23 & 2000 October 26 & 6663 & 1800 $\times$ 2 & 2\farcs33 & on-band \\
& & 6548, 6751 & 1800 $\times$ 2 & & off-band \\
III~Zw~35 & 2000 January 4 & 6746 & 1800 $\times$ 2 & 1\farcs62 & on-band \\
& & 6631, 6836 & 1800 $\times$ 2 & & off-band \\
NGC~695 & 2000 January 4 & 6773 & 1800 $\times$ 2 & 1\farcs70 & on-band \\
& & 6657 & 1800 & & off-band \\
NGC~828 & 2000 October 27 & 6681 & 1800 $+$ 747 & 2\farcs42 & on-band \\
& & 6769 & 1800 & & off-band \\
NGC~834 & 2000 October 30 & 6663 & 1800 $+$ 1389 & 2\farcs01 & on-band \\
& & 6752 & 1800 & & off-band \\
NGC~877 & 2000 October 26 & 6648 & 1800 & 2\farcs17 & on-band \\
& & 6534, 6736 & 1800 $\times$ 2 & & off-band \\
NGC~958 & 2000 January 8 & 6689 & 1800 & 1\farcs57 & on-band \\
& & 6573 & 1547 & & off-band \\
UGC~2238 & 1999 December 12 & 6703 & 1800 & 1\farcs78 & on-band \\
& & 6790 & 1800 & & off-band \\
NGC~1614 & 1999 December 12 & 6667 & 900 $\times$ 2 & 1\farcs83 & on-band \\
& & 6756 & 1800 & & off-band \\
NGC~2623 & 2000 October 30 & 6684 & 1800 & 2\farcs05 & on-band \\
& & 6773 & 1800 & & off-band \\
Arp~55 & 2000 January 8 & 6821 & 1800 $\times$ 2 & 1\farcs67 & on-band \\
& & 6705, 6920 & 1800 $\times$ 2 & & off-band \\
NGC~3110 & 2000 January 8 & 6673 & 1800 & 1\farcs35 & on-band \\
& & 6558, 6762 & 1800 $\times$ 2 & & off-band \\
Arp~148 & 2000 January 4 & 6789 & 1800 $\times$ 2 & 1\farcs73 & on-band \\
& & 6672, 6860 & 1800 $\times$ 2 & & off-band \\
IC~2810 & 2000 January 4 & 6647 & 1800 $\times$ 2 & 2\farcs05 & on-band \\
& & 6672, 6860 & 1800 $\times$ 2 & & off-band \\
NGC~4194 & 2001 April 19 & 6618 & 1800 $\times$ 2 & 1\farcs75 & on-band \\
& & 6504, 6706 & 1800 $\times$ 2 & & off-band \\
UGC~8335 & 2001 April 22 & 6766 & 1800 $\times$ 2 & 2\farcs28 & on-band \\
& & 6650, 6856 & 1800 $\times$ 2 & & off-band \\
NGC~5653 & 2001 April 22 & 6641 & 1800 & 2\farcs01 & on-band \\
& & 6527, 6729 & 1800 $\times$ 2 & & off-band \\
Zw~049.057 & 2001 April 22 & 6641 & 1800 & 2\farcs83 & on-band \\
& & 6527 & 1800 & & off-band \\
NGC~6090 & 2001 April 19 & 6754 & 1800 & 1\farcs80 & on-band \\
& & 6638 & 1800 & & off-band \\
NGC~6621/22 & 2001 April 22 & 6696 & 1800 $\times$ 2 & 1\farcs49 & on-band \\
& & 6581, 6786 & 1800 $\times$ 2 & & off-band \\
NGC~7771 & 2000 October 30 & 6657 & 1800 $\times$ 2 & 2\farcs52 & on-band \\
& & 6542, 6746 & 1800 $\times$ 2 & & off-band \\
Mrk~331 & 2000 October 27 & 6683 & 1800 $\times$ 2 & 1\farcs89 & on-band \\
& & 6568, 6772 & 1800 $\times$ 2 & & off-band \\
\enddata

\tablenotetext{a}{Estimated from FWHMs of field stars in the final images after data reduction.}

\end{deluxetable}

\clearpage
\begin{deluxetable}{lccc}
\tablewidth{0pt}
\tablecaption{Predicted infrared luminosities for paired galaxies from
radio continuum data.\label{pair_list}}
\tablehead{
\colhead{Name} & \colhead{$L_{\rm IR}$} & &
\colhead{Reference} \\
& \colhead{total} & \colhead{expected} & 
}
\startdata
III Zw 35 & 11.55 & & 1 \\
\multicolumn{1}{r}{north} & & 11.55 \\
\multicolumn{1}{r}{south} & & --- \\
Arp 55 & 11.70 & & 2 \\
\multicolumn{1}{r}{east} & & 11.55 \\
\multicolumn{1}{r}{west} & & 11.16 \\
UGC 8335\hspace{1em} & 11.76 & & 2 \\
\multicolumn{1}{r}{southeast} & & 11.68 \\
\multicolumn{1}{r}{northwest} & & 10.88 \\
NGC 6090 & 11.51 & & 3 \\
\multicolumn{1}{r}{northeast} & & 11.47 \\
\multicolumn{1}{r}{southwest} & & 10.50 \\
NGC 6621/22 & 11.22 & & 4 \\
\multicolumn{1}{r}{NGC 6621} & & 11.22 \\
\multicolumn{1}{r}{NGC 6622} & & --- \\
NGC 877 & 11.03 & 10.95 & 2 \\
NGC 3110 & 11.26 & 11.23 & 2 \\
IC 2810 & 11.53 & 11.36 & 2 \\
NGC 7771 & 11.35 & 11.29 & 2 \\
Mrk 331 & 11.41 & 11.41 & 5 \\
\enddata
\tablerefs{
(1) \citealt{chapman90}; (2) \citealt{condon90};
(3) \citealt{hummel87}; (4) \citealt{condon96};
(5) \citealt{zink00}.
}
\end{deluxetable}

\clearpage
\begin{deluxetable}{lccccccccc}
\tabletypesize{\footnotesize}
\rotate
\tablewidth{0pt}
\tablecaption{The indicators of H$\alpha$ distribution and global properties of
individual objects. \label{global_prop}}
\tablehead{
\colhead{Name} & \colhead{Morphological} & \colhead{Type} &
\colhead{$\log P/T$} & \colhead{$\log S_{\rm H\alpha}$} &
\colhead{$\log (L_{\rm IR}/M_{\rm H_2})$} &
\colhead{$\log (f_{60}/f_{25})$} & \colhead{$\log (f_{100}/f_{60})$} &
\colhead{$C_{60\mu}$\tablenotemark{b}} &
\colhead{$\log (L_{\rm IR}/L_{\rm H\alpha})$} \\
& \colhead{Class\tablenotemark{a}}
& & \colhead{(kpc$^{-2}$)} & \colhead{(kpc$^2$)} &
\colhead{($L_\odot$)} & \colhead{($L_\odot/M_\odot$)} & &
}
\startdata
NGC~23 & S & 3 & -0.879 & 2.036 & 1.37 & 0.85 & 0.23 & 1.19 & 2.927 \\
III~Zw~35 & C & 1 & -0.720 & 1.006 & 1.71 & 1.07 & 0.06 & 2.54 & 4.123 \\
NGC~695 & S & 4 & -1.685 & 2.313 & 1.29 & 0.97 & 0.26 & 1.39 & 2.922 \\
NGC~828 & S & 4 & -1.377 & \nodata & 1.18 & 1.06 & 0.31 & 1.25 & \nodata \\
NGC~834 & S & 4 & -1.125 & 1.901 & 1.39 & 0.89 & 0.27 & 1.15 & 2.958 \\
NGC~877 & P & 4 & -1.771 & 2.581 & 1.40 & 0.92 & 0.30 & 1.09 & 2.923 \\
NGC~958 & S & 4 & -1.887 & 2.608 & 1.02 & 0.79 & 0.40 & 0.26 & 2.844 \\
UGC~2238 & S & 3 & -1.357 & 1.783 & 1.13 & 1.11 & 0.27 & 1.71 & 3.607 \\
NGC~1614 & M & 2 & -0.602 & 2.122 & 1.58 & 0.63 & 0.04 & 1.48 & 3.085 \\
NGC~2623 & M & 1 & -0.746 & 1.211 & 1.77 & 1.14 & 0.03 & 2.89 & 4.557 \\
Arp~55 & C & 3 & -1.105 & 1.373 & 1.24 & 1.02 & 0.19 & 1.81 & 3.692 \\
NGC~3110 & P & 4 & -1.442 & 2.292 & 0.96 & 1.03 & 0.30 & 1.36 & 2.853 \\
Arp~148 & M & \nodata & -1.446 & 1.781 & 1.41 & 1.27 & 0.20 & 2.45 & 3.613 \\
IC~2810 & P & 3 & -1.057 & 1.532 & 1.27 & 0.98 & 0.24 & 1.49 & 3.736 \\
NGC~4194 & M & 2 & -0.577 & 1.939 & 1.81 & 0.75 & 0.01 & 1.93 & 2.954 \\
UGC~8335 & C & 2 & -0.660 & 1.624 & 1.63 & 0.79 & 0.03 & 1.93 & 3.341 \\
NGC~5653 & S & 4 & -1.008 & 1.965 & 1.40 & 0.89 & 0.33 & 0.86 & 2.827 \\
Zw~049.057 & S & 1 & -0.444 & 0.975 & 1.75 & 1.35 & 0.15 & 2.88 & 4.741 \\
NGC~6090 & C & 2 & -0.840 & \nodata & 1.36 & 0.71 & 0.17 & 1.08 & \nodata \\
NGC~6621/22 & C & 3 & -0.877 & 1.861 & 1.21 & 0.84 & 0.25 & 1.04 & 3.202 \\
NGC~7771 & P & 3 & -0.914 & 2.101 & 1.40 & 0.97 & 0.26 & 1.38 & 3.541 \\
Mrk~331 & P & 2 & -0.640 & \nodata & 1.30 & 0.83 & 0.08 & 1.82 & \nodata \\
\enddata

\tablenotetext{a}{M=merger, C=close pair, P=pair, and S=single}
\tablenotetext{b}{$C_{60\mu} \equiv \alpha_{25\mu,60\mu} -
\alpha_{60\mu,100\mu}$}

\end{deluxetable}

\clearpage
\begin{deluxetable}{lcccccccccccccccccc}
\tabletypesize{\footnotesize}
\rotate
\tablewidth{0pt}
\tablecaption{Mean and standard deviation of star-forming properties
for each type of object.\label{global_differ_table}}
\tablehead{
\colhead{Type} & \colhead{$P/T$} & & \colhead{$S_{\rm H\alpha}$} & &
\colhead{$L_{\rm IR}$} & & \colhead{$L_{\rm IR}/M_{\rm H_2}$} & &
\colhead{$f_{60}/f_{25}$} & & \colhead{$f_{100}/f_{60}$} & &
\colhead{$L_{\rm IR}/L_{\rm H\alpha}$} & & \colhead{$C_{60\mu}$} & \\
& \colhead{(kpc$^{-2}$)} & & \colhead{(kpc$^2$)} &
& \colhead{($L_\odot$}) & & \colhead{($L_\odot/M_\odot$)} & \\
& \colhead{mean} & \colhead{$\sigma$} & \colhead{mean} & \colhead{$\sigma$}
& \colhead{mean} & \colhead{$\sigma$} & \colhead{mean} & \colhead{$\sigma$}
& \colhead{mean} & \colhead{$\sigma$} & \colhead{mean} & \colhead{$\sigma$}
& \colhead{mean} & \colhead{$\sigma$} & \colhead{mean} & \colhead{$\sigma$} \\
}
\startdata
1 & -0.637 & 0.137 & 1.064 & 0.105 & 11.437 & 0.153 & 1.743 & 0.025 &
1.191 & 0.120 & 0.081 & 0.052 & 4.474 & 0.259 & 2.767 & 0.163 \\
2 & -0.664 & 0.093 & 1.895 & 0.206 & 11.476 & 0.225 & 1.536 & 0.186 &
0.741 & 0.070 & 0.067 & 0.059 & 3.127 & 0.161 & 1.646 & 0.329 \\
3 & -1.032 & 0.170 & 1.781 & 0.259 & 11.353 & 0.211 & 1.270 & 0.092 &
0.963 & 0.095 & 0.244 & 0.026 & 3.451 & 0.291 & 1.436 & 0.271 \\
4 & -1.471 & 0.305 & 2.277 & 0.271 & 11.196 & 0.217 & 1.234 & 0.172 &
0.938 & 0.076 & 0.314 & 0.043 & 2.888 & 0.049 & 1.050 & 0.363 \\
\enddata
\tablecomments{All values except for $C_{60\mu}$ use logarithmic scales.}
\end{deluxetable}

\appendix

\section{NGC~23 (Mrk~545)}
\label{section_n23}

NGC~23 (Figure~1a) is a barred spiral galaxy paired with NGC~26,
a spiral galaxy, with an apparent separation of $\sim$ 9\farcm2 ($\sim$
160 kpc).
In the H$\alpha$ image, NGC~23 shows a bright nuclear component
and extended emission having the appearance of an integral sign.
The nuclear-component has an asymmetric shape with a peak
position in H$\alpha$ offset by 1\farcs7 to the west of the nucleus.
In the H$\alpha$ equivalent-width map,
a ring-like feature that surrounds the nucleus at a radius of $\sim$
7\arcsec \ ($\sim$ 2 kpc) is seen.
This means that the H$\alpha$ distribution is less sharply peaked than
the continuum in the nuclear region.

The continuum source located $\sim$ 30\arcsec \ southeast of the
nucleus is a foreground star.

\section{III~Zw~35}
\label{section_zw35}

III~Zw~35 (Figure~1b) is composed of two galaxies separated by $\sim$
9\arcsec \ ($\sim$ 5 kpc) along a position angle of 20$^\circ$.
The northern galaxy is the dominant source in the optical continuum
image.
In H$\alpha$, it appears compact but slightly elongated 
in the direction of the minor axis.
The southern galaxy also shows clear elongation along a P.A. of
$\sim$ 23$^\circ$.
Although the two galaxies have comparable fluxes in H$\alpha$,
the majority of the radio continuum (and also probably FIR) emission
originates in the northern galaxy \citep{chapman90}.

From optical emission lines, the nucleus of the northern galaxy is
classified as a LINER and the southern one as an \ion{H}{2} nucleus
\citep{veilleux95,chapman90}.
The morphological appearance in radio continuum emission
suggests that star-forming activity dominates the energetics of
the northern galaxy \citep{pihlstrom01}; the classification as
a LINER may be due to a contribution from shock heating, possibly
driven by superwind activity \citep{taniguchi99,lutz99}.
The elongated appearance described above is consistent with this
interpretation: a significant [\ion{N}{2}] contribution to the
H$\alpha$ emission may enhance the shock heated regions, which are extended
in the direction of the minor axis.

Note that the analyses presented in \S~\ref{section_ha_con_diagram} are
made only for the northern galaxy.

\section{NGC~695}
\label{section_n695}

In the optical/near-IR continuum images, NGC~695 (Figure~1c) has the appearance of a disturbed face-on spiral galaxy
\citep{hutchings91,smith96}.
Our H$\alpha$ image reveals more clearly the knotty and chaotic
distribution of star-forming regions in this galaxy.
The brightest source in H$\alpha$ is located 5\farcs6 (3.5 kpc)
north-west of the nucleus, corresponding to the position where the inner spiral arm seen in the continuum
image terminates.

\section{NGC~828}
\label{section_n828}

NGC~828 (Figure~1d) is a disturbed spiral galaxy with a prominent dust
lane to the southwest side of the main body.
In the H$\alpha$ image, a diffuse source at the center and
two sources located nearly symmetrically along the major axis of the
galaxy can be seen.
The surface brightnesses of the three sources are comparable.

\citet{wang91} suggested, based on an anomaly in the CO rotation curve and the presence of two emission peaks (the nucleus and one $\sim$ 4\arcsec \ to the southeast) in the radio continuum image, that this galaxy
is an ongoing merger.

\section{NGC~834}
\label{section_n834}

NGC~834 (Figure~1e) is a spiral galaxy with a complicated appearance.
It belongs to a group \citep{garcia93}, and its nearest neighbors are
NGC~841, $\sim$ 10\farcm6 ($\sim$ 190 kpc) to the south, and UGC~1673, $\sim$ 9\farcm4 ($\sim$ 170 kpc) to the north.
The brightest H$\alpha$ source lies $\sim$ 8\arcsec \ (2.4 kpc) northwest
of the nucleus. A large ring of star-forming regions with a radius of $\sim$
15\arcsec \ is seen in the H$\alpha$ equivalent-width map.
This ring is also seen in the combined color-index ($Q_{BVI}$)
image of \citet{bizyaev01}.

\section{NGC~877}
\label{section_n877}

NGC~877 (Figure~1f) is a late-type spiral galaxy paired with NGC~876,
a spiral galaxy 2\farcm1 ($\sim$ 31 kpc) to the southwest.
These two galaxies are also members of a group that includes NGC~870
and NGC~871.

The H$\alpha$ image of NGC~877 shows numerous knots distributed along
the two spiral arms, which appear slightly disturbed.
It is interesting that there are several
pairs of relatively bright H$\alpha$ sources located at nearly opposite
positions with respect to the nucleus; this is a phenomenon seen in several
grand-design spirals without bars \citep{rozas98}.
The pair of brightest sources
located at the ends of the two spiral arms is a typical example.
The H$\alpha$ emission from the nucleus is much fainter than that from the spiral
arms.

\section{NGC~958}
\label{section_n958}

NGC~958 (Figure~1g) is a late-type spiral galaxy viewed at a high inclination.
There are several faint galaxy-like objects around NGC~958.
As in normal late-type spirals, the H$\alpha$ emission is distributed
along the spiral arms, and the nucleus is not detected in H$\alpha$.

\section{UGC~2238}
\label{section_u2238}

UGC~2238 (Figure~1h) has a highly elongated appearance.
In deep optical/near-infrared images it has two faint spiral arms and a faint tail-like structure.
\citet{smith96} suggested, based on the tail-like structure and the single nucleus, that this galaxy is an advanced merger.
The H$\alpha$ image shows that UGC~2238
has a relatively flat distribution of H$\alpha$ emission as compared to the other
galaxies in our sample. The nucleus is located at the southwest edge of
the emission-line region.

\section{NGC~1614}
\label{section_n1614}

NGC~1614 (Figure~1i) is one of the most well-studied objects in our sample.
It is a strongly interacting galaxy in a late stage of the merging process,
with spectacular tidal features.
Our H$\alpha$ image has identical features to the
H$\alpha+$[\ion{N}{2}] image by AHM and \citet{dopita02}:
a bright nuclear source, the inner spiral
arms, and \ion{H}{2} regions along the southwestern linear tail. The
nuclear source is very compact and the nuclear ring \citep{alonso01} is
not spatially
resolved in our image [with $\sim$ 1\farcs8 ($\sim$ 0.5 kpc) seeing].

\section{NGC~2623}
\label{section_n2623}

NGC~2623 (Figure~1j) is a nearly completed merger showing an $r^{1/4}$ profile
at K-band
\citep{wright90,stanford91,chitre02}
and two long tidal tails. Extensive observations of this galaxy have been made
from the X-ray to the radio. The radio continuum \citep{condon91},
mid-infrared \citep{soifer01}, CO \citep{bryant99},
and H$\alpha$+[\ion{N}{2}] (AHM) emission distributions are dominated
by a nuclear compact source.
In our H$\alpha$ image, the nucleus also dominates the emission, and
there is little emission in the outer regions, except for
a faint source $\sim$ 8\arcsec \ ($\sim$ 3 kpc) southeast
of the nucleus.

\section{Arp~55 (UGC~4881)}
\label{section_a55}

Arp~55 (Figure~1k) is an interacting system of two galaxies with a separation
of $\sim$ 11\farcs5 ($\sim$ 9 kpc).
To the east of the main body, a bright tail extends towards the south.
The western nucleus shows compact H$\alpha$ emission with a slight extension
in the north-south direction. The eastern
nucleus shows an arm-like structure extending towards the north, and the peak
position in H$\alpha$ is
displaced by $\sim$ 1\arcsec \ to the southwest of the nucleus.
There are also four H$\alpha$ sources along the eastern tail.

Most of the radio continuum emission originates in the eastern galaxy
\citep{condon90}, and the analyses presented in
\S~\ref{section_ha_con_diagram} are made only for the eastern galaxy.

\section{NGC~3110}
\label{section_n3110}

NGC~3110 (Figure~1l) is a distorted spiral galaxy with a small companion
$\sim$ 1\farcm8 ($\sim$ 36 kpc) to the southwest (MCG~--01-26-013).
An H$\alpha$+[\ion{N}{2}]
image of this galaxy \citep{dopita02} shows bright H$\alpha$ knots
along the two spiral arms.
Our H$\alpha$ image also shows the knots, as well as a possible bar-like
structure through the nucleus with P.A. $\sim$ 60$^\circ$ and length
$\sim$ 20\arcsec.
The bar-like structure cannot be seen in our continuum image.
\citet{zink00} investigated the distribution of 100\micron \ emission and
concluded that the disk or arms of NGC~3110 contribute up to 35\% of the
total FIR flux.

The companion galaxy MCG~--01-26-013 (Figure~2a)
shows H$\alpha$ emission associated with the nucleus \citep{dopita02}.
There is also an H$\alpha$ source $\sim$ 3\arcsec \ east of the nucleus.

\section{Arp~148}
\label{section_a148}

Arp~148 (Figure~1m), known as Mayall's Object, is a ring galaxy composed of
an elongated main body with double components separated by $\sim$ 5\arcsec
\ ($\sim$ 4 kpc) and an oval ring with a diameter of $\sim$ 16\arcsec
\ to the west of the main body.
From JHKL-band
mapping, \citet{joy87} found that the nucleus of Arp~148 lies between
the double optical components of the main body, and concluded that the nucleus
suffers from heavy extinction. Despite the heavy extinction,
the H$\alpha$ emission peaks midway between the double
components.
The ring is more prominent in H$\alpha$ than continuum light.

\section{IC~2810}
\label{section_ic2810}

IC~2810 (Figure~1n) is a disk galaxy viewed nearly edge-on.
In contrast to the continuum image, which shows a bright nucleus and a linearly
extended disk, the H$\alpha$ image reveals a warped structure that is
more distinct on the south side of the nucleus.
The brightest H$\alpha$ emission in the nucleus is extended by $\sim$
3\arcsec \ ($\sim$ 2 kpc) towards the north.

There is a small companion galaxy (IC~2810b), which is also
edge-on $\sim$ 70\arcsec \ ($\sim$ 48 kpc) to the southeast.
Compact H$\alpha$ emission is detected from the nucleus
(Figure~\ref{companion_image}b).

\section{NGC~4194}
\label{section_n4194}

NGC~4194 (Figure~1o), known as The Medusa, is a merger with tidal
features of peculiar appearance.
An H$\alpha$+[\ion{N}{2}] image is presented by AHM and
shows a single dominant nucleus with several circumnuclear hotspots
embedded in a faint halo of 3~kpc diameter.
Our H$\alpha$ image reproduces these features, though with coarser spatial
resolution.
A close inspection of the H$\alpha$ image suggests that the nuclear source has an arc-like appearance. This arc-like structure is composed of the brightest
nuclear source and two extensions ($\sim$ 3\arcsec, $\sim$ 0.6 kpc)
to the west and the south. The western
extension is reminiscent of the one seen in the radio continuum map by
\citet[see also \citealt{aalto00}]{condon90}.

\section{UGC~8335 (Arp~238)}
\label{section_u8335}

UGC~8335 (Figure~1p) is a pair of spiral galaxies separated by 35\arcsec
\ ($\sim$ 22 kpc), with a linking bridge and strong tidal tails.
Most of the H$\alpha$ flux is emitted by a compact source at the
nucleus of the eastern galaxy.
This is also true in radio continuum emission and probably in FIR
(see Table~\ref{pair_list}).
The western galaxy has a relatively extended distribution of H$\alpha$
emission, with an elongated nuclear component and several sources
along the western tidal tail.

Note that the analyses presented in \S~\ref{section_ha_con_diagram} are
made only for the eastern galaxy.

\section{NGC~5653}
\label{section_n5653}

NGC~5653 (Figure~1q) is an isolated spiral galaxy
classified as a lopsided galaxy (Rudnick et al. 2000).
The H$\alpha$ emission is distributed along the spiral arms and the
brightest \ion{H}{2} regions are located $\sim$ 10\arcsec
\ ($\sim$ 2.6 kpc) west of the nucleus.
The Pa$\alpha$ distribution in NGC~5653 derived from HST/NICMOS observations
\citep{alonso02} shows nearly identical features to our H$\alpha$
image: bright \ion{H}{2} regions to the west of the nucleus and no
bright nuclear emission.

\section{Zw~049.057}
\label{section_zw49}

Zw~049.057 (Figure~1r) is a highly inclined disk galaxy.
Although several galaxies are seen around Zw~049.057,
those with known redshifts belong to the background cluster of galaxies
Abell~2040 at a redshift of 0.046 and most of the others may also
belong to it.

The nucleus of Zw~049.057 suffers from heavy extinction
\citep{scoville00} and shows OH megamaser emission
\citep{baan87,martin88}.
The H$\alpha$ emission is very faint and confined to the nucleus.
A Pa$\alpha$ image of Zw~049.057 was presented by \citet{alonso02}
It shows only faint Pa$\alpha$ emission in the nuclear region.
In spite of its highly elongated appearance in continuum light,
\citet{scoville00} found that the near-infrared light profiles
are better fitted by an $r^{1/4}$ law than by an exponential disk profile.

\section{NGC~6090 (Mrk~496)}
\label{section_n6090}

NGC~6090 (Figure~1s) is a pair of spiral galaxies with a compact main body
and long
tidal tails \citep{bergvall81}. Although the two galaxies show strong H$\alpha$
emission at their nuclei, the H$\alpha$ emission appears to be
brighter on the companion-facing side of each galaxy. This effect is seen
more clearly in the equivalent-width map: the H$\alpha$ equivalent width
peaks 1\farcs5 ($\sim$ 1 kpc) southwest of the nucleus in the northeastern
galaxy and 1\farcs5 northeast in the southwestern galaxy.

Most of the radio continuum emission originates in the northeastern galaxy
\citep{hummel87}, and the analyses presented in
\S~\ref{section_ha_con_diagram} are made only for the northeastern galaxy.

\section{NGC~6621/22 (Arp~81)}
\label{section_n6621}

NGC~6621/22 (Figure~1t) is a pair of spiral galaxies and is included in the
"Toomre Sequence" (Toomre 1977) as an intermediate-stage merger.
The northwestern galaxy (NGC~6621)
contributes nearly all of the radio and FIR flux
\citep{condon96,bushouse98}.
The H$\alpha$ flux is also dominated by NGC~6621.
Excess H$\alpha$ emission is seen
from the overlapping regions of the two galaxies \citep[see also][]{xu00}.
The nuclear region of NGC~6621 shows two bright H$\alpha$ sources separated by
$\sim$ 3\arcsec \ ($\sim$ 1.3 kpc).
The southwestern source is the brightest in the entire system,
and the northeastern one coincides with the nucleus of NGC~6621.
We detected no evidence for H$\alpha$
emission from the nucleus of NGC~6622, consistent with the nuclear
spectrum of \citet{liu95}.
Faint H$\alpha$ emission in the tidal tail is seen $\sim$ 40\arcsec \ east
and $\sim$ 5\arcsec \ south of the NGC~6621 nucleus.

The analyses presented in \S~\ref{section_ha_con_diagram} are made only for
the eastern galaxy.

\section{NGC~7771}
\label{section_n7771}

NGC~7771 (Figure~1u) is a highly inclined, barred spiral galaxy in
a group with NGC~7769, NGC~7770 and NGC~7771A \citep{nordgren97},
and the latter two are also included in our field-of-view at distances of
$\sim$ 1\arcmin\ ($\sim$ 18~kpc) and $\sim$ 2\farcm8 ($\sim$ 48~kpc),
respectively.
While the CO flux reported in SSS refers only to NGC~7771, part of
the FIR emission measured by $IRAS$ may originate from NGC~7770
\citep{condon90}.
The nuclear region of NGC~7771 shows a complicated structure both in the
H$\alpha$ and continuum images, which may be due to the starburst ring,
which is thought to have a complicated star-formation history and extinction
\citep{smith99,reunanen00}.
The H$\alpha$ peak is offset by $\sim$ 2\farcs4 ($\sim$ 0.7 kpc) to the west
of the continuum peak.
This is reminiscent of the relation between the Br$\gamma$ and K-band
images \citep{reunanen00}.
H$\alpha$ emission is also detected along the bar and spiral arms.

NGC~7770 (Figure~2c) is a strong H$\alpha$ source, showing double peaks
in both H$\alpha$ and continuum emission, and has a comparable flux to NGC~7771.
The western peak in H$\alpha$ is coincident with the continuum peak, while
the eastern one is offset by $\sim$ 2\arcsec \ to the north of the eastern
continuum peak.
In addition, there is a tail-like structure extending to the west from the
northern part of the main body. This structure cannot be seen in our
continuum image, nor in the $I$-band image shown in \citet{smith99}.
We also detected faint H$\alpha$ emission from NGC~7771A (not shown).

\section{Mrk~331}
\label{section_m331}

Mrk~331 (Figure~1v) has two faint companions at $\sim$ 2\arcmin
\ ($\sim$ 42 kpc) to the west (UGC~12812) and $\sim$ 1\farcm4 ($\sim$ 30 kpc)
to the southwest.
The H$\alpha$ and continuum images show a strong peak at the nucleus.
While little structure can be seen in the circumnuclear region in either
 H$\alpha$ or continuum emission, the
equivalent-width map of Mrk~331 reveals a ring-shaped structure surrounding
the nucleus with a diameter of $\sim$ 5\arcsec \ ($\sim$ 1.8 kpc).
As Mrk~331 is known from high-spatial-resolution radio and MIR maps \citep[C91;][]{soifer01} to have a circumnuclear star-forming ring with
dimensions of 3\arcsec \ $\times$ 2\arcsec \ ,
this star-forming ring, when affected by the $\sim$ 2\arcsec \ seeing, may appear
as the larger ring in the equivalent-width map.

We also detected H$\alpha$ emission from both companion galaxies.
Both show extended and complex distributions with large
equivalent width (Figure~\ref{companion_image}d).

\end{document}